\documentclass[showpacs,aps,preprint]{revtex4}
\usepackage{graphicx}
\usepackage{amsmath,amssymb}

\def\gtsim{\mathrel{\hbox{\raise0.2ex
  \hbox{$>$}\kern-0.75em\raise-0.9ex\hbox{$\sim$}}}}
\def\ltsim{\mathrel{\hbox{\raise0.2ex
  \hbox{$<$}\kern-0.75em\raise-0.9ex\hbox{$\sim$}}}}

\newcommand{\bm}[1]{\mbox{\boldmath $#1$}} 

\begin{document}

\title{
Infinitesimal cranking for triaxial angular-momentum-projected
configuration-mixing calculation and its application to
the gamma vibrational band
}

\author{Shingo Tagami and Yoshifumi R. Shimizu}
\affiliation{Department of Physics, Graduate School of Science,
Kyushu University, Fukuoka 819-0395, Japan}


\begin{abstract}

Inclusion of time-odd components into the wave function
is important for reliable description of rotational motion
by the angular-momentum-projection method;
the cranking procedure with infinitesimal rotational frequency
is an efficient way to realize it.   In the present work
we investigate the effect of this infinitesimal cranking
for triaxially deformed nucleus, where there are three independent
cranking axes.  It is found that the effects of
cranking about three axes on the triaxial energy spectrum are
quite different and inclusion of all of them considerably modify
the resultant spectrum from the one obtained without cranking.
Employing the Gogny D1S force as an effective interaction,
we apply the method to the calculation of the multiple gamma vibrational bands
in $^{164}$Er as a typical example, where the angular-momentum-projected
configuration-mixing with respect to the triaxial shape degree of freedom
is performed.
With this method, both the $K=0$ and $K=4$ two-phonon gamma vibrational
bands are obtained with considerable anharmonicity.
Reasonably good agreement, though not perfect, is obtained
for both the spectrum and transition probabilities
with rather small average triaxial deformation
$\gamma\approx 9^\circ$ for the ground state rotational band.
The relation to the wobbling motion at high-spin states
is also briefly discussed.

\end{abstract}


\maketitle

\section{Introduction}
\label{sec:intro}

The angular-momentum-projection method is a fully microscopic means
to recover the rotational invariance, which is broken in the selfconsistently
determined nuclear mean-field, e.g.,
by the Hartree-Fock-Bogoliubov (HFB) calculation.
Although nice rotational spectrum is obtained by the projection
from one intrinsic mean-field,
it often happens that the moment of inertia is smaller in comparison
with the experimental data as long as the projection is performed from
the time-reversal invariant mean-field state.
Inclusion of the time-odd components is important for realistic
description of nuclear rotational motion, and one of the efficient ways
to realize it is the so-called cranking procedure,
which is justified by the variational point of view~\cite{RS80}.
Recently we have shown~\cite{TS12} that indeed the calculated moment of inertia
is considerably increased if the projection is performed from the cranked
mean-field state with very small cranking frequency.
Moreover, the resultant spectrum is independent of actual values of
the frequency if it is small enough~\cite{TS12}:
We call this procedure as ``infinitesimal cranking''.
In our previous studies the selfconsistent mean-field state
before the cranking is either axially symmetric~\cite{TS12} or
tetrahedrally-symmetric~\cite{TSD13,TSD15},
so that the direction of cranking axis does not matter;
there is only one rotational axis for the quantum mechanical
axially symmetric system, and the tetrahedral deformation is ``spherical''
in the sense that all rotational axes are equivalent, which is also
confirmed numerically~\cite{TSD13}.
In the present work we consider the case of the triaxial deformation,
where one can crank the mean-field state
around three independent rotational axes.
We study the effects of infinitesimal cranking around
three principal axes on the spectrum of triaxially deformed nucleus
obtained by the angular-momentum-projection method.

The second purpose of the present investigation is the description
of the gamma vibration by the angular-momentum-projection method.
The motivation emerged from the precedent analysis
by the so-called triaxial projected shell model
approach~\cite{SHS00,SBS08}:  It is concluded that rather large
triaxial deformation of the mean-field is necessary to reproduce
the very low-lying nature of the gamma vibration,
see also Refs.~\cite{HS95,VHS98}
for studies with the axial projected shell model.
The gamma vibration is the most well recognized
collective vibration in atomic nuclei~\cite{EG70,BM75,RS80}.
In the rare earth region the ground state is believed to be axially deformed
according to the mean-field calculations by,
e.g., the Strutinsky shell correction method and/or the Skyrme HFB method,
and the gamma vibration is interpreted as a surface vibration that
dynamically breaks the axial symmetry.
In fact the gamma vibration has been studied
by the random phase approximation (RPA) calculation
with the schematic $QQ$ type interaction
based on the axially symmetric vacuum state~\cite{BFM65},
although the strong anharmonicity for two-phonon states exists;
see, e.g., Refs.~\cite{DH82,MM85,MM86,MM87} and references therein.
However, there is a long history on the interpretation of the gamma vibration;
by employing the asymmetric (triaxial) rotor model~\cite{DF58}
it was discussed in the early days that
its low-lying nature indicates the considerable triaxial deformation,
although it is very difficult to draw a definite conclusion
from the existing experimental data.
Therefore the analysis by the triaxial projected shell model~\cite{SHS00,SBS08}
revived the old problem; whether the nucleus in the rare earth region
is axially symmetric or triaxially deformed.
It is worthwhile mentioning that in the triaxially deformed case
the high-spin part of the multiple rotational bands
based on the excitations of the gamma vibration
is interpreted as the wobbling-phonon bands~\cite{BM75}.
We briefly discuss also this interesting issue of the relation
between the multiple gamma bands and the wobbling band.

As for the proper treatment of the triaxial degree of freedom,
we use the configuration-mixing,
or the generator coordinate method (GCM)~\cite{RS80},
on top of the angular-momentum-projection.
One of the great merits of this microscopic approach is
that not only the energy spectrum but also the transition probability
can be calculated full quantum mechanically without any ambiguity.
It is known that the $E2$ transition probability between the ground state
and the gamma vibrational state is overestimated in the RPA approach
with schematic interaction by a factor $3-4$ in the rare earth region,
if the Nilsson potential is used as a mean-field~\cite{DH82};
the situation is improved if the Woods-Saxon potential is used instead
but the $B(E2)$ is still overestimated by a factor $2-3$~\cite{Shoji09}.
It will be shown that this problem is greatly improved
in our angular-momentum-projected configuration-mixing approach.
We employ the Gogny D1S force~\cite{D1S} as an effective interaction,
and select the nucleus $^{164}$Er as a typical example of rare earth nuclei.
We present and discuss our results in comparison with experimental data
and with the previous pioneering works~\cite{SHS00}.
The paper is organized as follows: The basic formulation of the method
employed is briefly outlined in Sec.~\ref{sec:theorfr}.  The results
of the numerical calculations are presented in Sec.~\ref{sec:result},
where the effects of the triaxial deformation and the infinitesimal cranking
are discussed.
In the final section~\ref{sec:summary},
we give summary of the present work and further discussion.

\section{Theoretical framework}
\label{sec:theorfr}

\subsection{Angular-momentum-projected configuration-mixing}
\label{sec:ampcm}

The calculational method we employ is the standard one~\cite{RS80},
and the wave function $|\Psi^I_{M,\alpha}\rangle$, where $\alpha$ specifies
the quantum numbers other than the angular momentum $(IM)$,
is obtained in the form,
\begin{equation}
 |\Psi^I_{M,\alpha}\rangle = \sum_{Kn} g^I_{Kn,\alpha}\,
 P^I_{MK}|\Phi_n^{}\rangle,
\label{eq:proj}
\end{equation}
where the operator $P^I_{MK}$ is the angular-momentum-projector,
and $|\Phi_n^{}\rangle$ $(n=1,2,\cdots)$ are the mean-field states,
which are specified in more details in the following.
The amplitude $g^I_{Kn,\alpha}$ is determined by the so-called
Hill-Wheeler equation,
\begin{equation}
 \sum_{K'n'}{\cal H}^I_{K,K'n'}\ g^I_{K'n',\alpha} =
 E^I_\alpha\,
 \sum_{K'n'}{\cal N}^I_{Kn,K'n'}\ g^I_{K'n',\alpha},
\label{eq:HW}
\end{equation}
with definitions of the Hamiltonian and norm kernels,
\begin{equation}
 \left\{ \begin{array}{c}
   {\cal H}^I_{Kn,K'n'} \\ {\cal N}^I_{Kn,K'n'} \end{array}
 \right\} = \langle \Phi_n |
 \left\{ \begin{array}{c}
   H \\ 1 \end{array}
 \right\} P_{KK'}^I | \Phi_{n'} \rangle;
\label{eq:kernels}
\end{equation}
see e.g. Ref.~\cite{RS80} for more details.
We do not perform the number projection in the present work,
and treat the number conservation approximately
by replacing $H \rightarrow H-\lambda_\nu (N-N_0)-\lambda_\pi (Z-Z_0)$,
where $N_0$ and $Z_0$ are the neutron and proton numbers to be fixed.
As for the neutron and proton chemical potentials
$\lambda_\nu$ and $\lambda_\pi$
we use those obtained for the HFB ground state.

A set of the mean-field states, $|\Phi_n^{}\rangle$ $(n=1,2,\cdots)$,
are calculated by the constrained Hartree-Fock-Bogoliubov (HFB) method
with the quadrupole operators, $Q_{20}$ and $Q_{22}$
in $Q_{2m}\equiv r^2Y_{2m}$ with $Y_{lm}$ being the spherical harmonics,
as constraints.
In place of the HFB expectation values,
$\langle Q_{20} \rangle$ and $\langle Q_{22} \rangle$, we actually constrain
two quantities $(Q,\gamma)$ defined by
\begin{equation}
 Q \equiv \sqrt{\langle Q_{20} \rangle^2 + 2\langle Q_{22} \rangle^2},
 \qquad
 \gamma \equiv
   -\tan^{-1}{\biggl(\frac{\sqrt{2}{\langle Q_{22} \rangle}}
           {\langle Q_{20} \rangle}\biggr)},
\label{eq:Qgamma}
\end{equation}
which correspond to the quadrupole deformation parameters $(\beta_2,\gamma)$
with $\beta_2=\frac{4\pi}{5}Q/(A\langle r^2 \rangle)$
(note the Lund convention for the sign of $\gamma$).
We employ the augmented Lagrangian method in Ref.~\cite{SBN10}
in order to achieve strict fulfillment of the constraints
$(Q,\gamma)$ for arbitrarily desired values.
In the present work, we mainly keep the magnitude of
the quadrupole deformation $Q$ as
that of the ground state and vary only the value of $\gamma$.
Then the projected wave function is obtained as a function of $\gamma$,
\begin{equation}
 |\Psi^I_{M,\alpha}(\gamma)\rangle =
 \sum_{K} g^I_{K,\alpha}(\gamma)\,
 P^I_{MK}|\Phi(\gamma)\rangle.
\label{eq:GCMgamNocm}
\end{equation}
Instead of the amplitude $g^I_{K,\alpha}(\gamma)$ in Eq.~(\ref{eq:GCMgamNocm}),
the properly normalized amplitude~\cite{RS80} is necessary in some cases:
\begin{equation}
f^I_{K,\alpha}(\gamma)=\sum_{K'}
\bigl(\sqrt{{\cal N}^I}\,\bigr)_{K,K'}\, g^I_{K',\alpha}(\gamma),
\label{eq:normfNocm}
\end{equation}
where the quantity $\sqrt{{\cal N}^I}$ denotes
the square-root matrix of the norm kernel.

The triaxial deformation should be finally treated dynamically.
This is done by the configuration-mixing or
the generator coordinate method (GCM) with respect to the triaxiality $\gamma$.
Thus the wave function is obtained by
\begin{equation}
 |\Psi^I_{M,\alpha}\rangle =
 \int d\gamma\,\sum_{K} g^I_{K,\alpha}(\gamma)\,
 P^I_{MK}|\Phi(\gamma)\rangle,
\label{eq:GCMgam}
\end{equation}
in the continuum limit of the variable $\gamma$ under the fixed value of $Q$.
In this case the norm kernel is expressed like
${\cal N}^I_{K,K'}(\gamma,\gamma')$ and the properly normalized amplitude
is calculated by
\begin{equation}
f^I_{K,\alpha}(\gamma)=\int d\gamma'\,\sum_{K'}
\bigl(\sqrt{{\cal N}^I}\,\bigr)_{K,K'}(\gamma,\gamma')\,
g^I_{K',\alpha}(\gamma'),
\label{eq:normf}
\end{equation}
with which the probability distribution
of the eigenstate $|\Psi^I_{M,\alpha}\rangle$
with respect to the $\gamma$ coordinate,
\begin{equation}
p^I_{\alpha}(\gamma)=\sum_{K}|f^I_{K,\alpha}(\gamma)|^2,
\label{eq:probf}
\end{equation}
can be studied.

We have recently developed an efficient method to perform
the angular-momentum-projection calculation~\cite{TS12},
and it is successfully applied to the study of
the nuclear tetrahedral symmetry~\cite{TSD13,TSD15}:
The method is fully employed also in the present work.
The same configuration-mixing method has been also utilized for the study
of the rotational motion in Ref.~\cite{STS15}, where the generator coordinate
is chosen to be the rotational frequency $\omega_{\rm rot}$ instead of
the triaxiality parameter $\gamma$.   See Refs.~\cite{TS12,TSD15,STS15}
for more details of our method of calculation.

\subsection{Cranking procedure with infinitesimal rotational frequencies}
\label{sec:crank}

As has been stressed in Refs.~\cite{TS12,TSD13,STS15},
it is very important to include the time-odd components
in the mean-field wave function $|\Phi(\gamma)\rangle$,
from which the angular-momentum-projection is performed,
in order to properly describe the moment of inertia of rotational band.
This can be achieved by the cranking procedure
with infinitesimally small rotational frequencies.
Namely, considering that we are dealing with the triaxial deformation,
the ``3D cranked Hamiltonian''
with the angular momentum operators $\bm{J}\equiv(J_x,J_y,J_z)$,
\begin{equation}
 H'\equiv H-\bm{\omega}\cdot\bm{J}=H-\omega_x J_x - \omega_y J_y - \omega_z J_z,
\label{eq:Hcrank}
\end{equation}
is used in the constrained HFB calculation.
If the cranking axis specified by the direction of
the frequency vector $\bm{\omega}$ does not coincide with
one of the inertia axes, i.e., in the case of the tilted-axis cranking,
the intrinsic coordinate frame rotates
during the iterations of selfconsistent calculation.
Therefore we require the following additional principal-axis constraints
according to Ref.~\cite{KO81},
\begin{equation}
 \langle Q_{21} \rangle=\langle Q_{2-1} \rangle=0\quad\mbox{and}\quad
 \langle Q_{22} \rangle=\langle Q_{2-2} \rangle .
\label{eq:pacd}
\end{equation}
In fact the quantity $\langle Q_{22} \rangle$ takes real value
by the second condition in Eq.~(\ref{eq:pacd})
and then Eq.~(\ref{eq:Qgamma}) is well-defined.
In the present work we only consider the case of infinitesimal cranking, but
the large rotational frequencies with the tilted-axis cranking can be applied
to study various high-spin phenomena, see e.g. Ref.~\cite{Fra01}.

In this subsection we discuss what kind of time-odd components are
included by the 3D cranking in Eq.~(\ref{eq:Hcrank}),
and show that the result of angular-momentum-projection
neither depend on the particular choice of
the cranking frequency nor of the cranking axis as long as
the values of frequencies $\bm{\omega}=(\omega_x,\omega_y,\omega_z)$ are small.
Therefore there is no ambiguity in this infinitesimal cranking procedure.
The content was partly discussed in relation to the zero frequency limit
in Sec.IIIC of Ref.~\cite{TS12},
but the independence of the result was not fully explained.
Moreover, the case considered was somewhat specific;
the ground state before the cranking is axially symmetric
and its wave function has only $K=0$ components.
Here we consider more general cases for even-even nuclei,
where the ground mean-field state before the cranking is time-reversal invariant
and has the $D_2$-symmetry belonging to 
the totally symmetric $(r_x,r_y,r_z)=(+1,+1,+1)$ representation~\cite{BM75}.
Here $r_i$ $(i=x,y,z)$ is the quantum number of the $i$-signature,
i.e., the $\pi$-rotation around the $i$-axis, and $r_x r_y r_z=+1$.
The classification of the projected states according to
the $y$-signature quantum number is convenient for the projection calculation
using the usual rotation operator
$R(\alpha,\beta,\gamma)=e^{i\gamma J_z}e^{i\beta J_y}e^{i\alpha J_z}$.
In Ref.~\cite{TS12} the $x$-signature is employed instead
but the conclusion does not change.
For the notational simplicity we neglect the irrelevant configuration-mixing
in this subsection but the extension is trivial.
Then the wave function in Eq.~(\ref{eq:proj}) is written
in the $y$-signature classified way~\cite{BM75} as
\begin{equation}
 |\Psi^I_{M,\alpha}\rangle =
  \sum_{K\ge 0}\left(
   \tilde{g}^I_{K,\alpha}\, \tilde{P}^I_{MK}
  +\tilde{g}^I_{\bar{K},\alpha}\, \tilde{P}^I_{M\bar{K}} \right)
  |\Phi\rangle,
\label{eq:projRy}
\end{equation}
where the amplitudes
$(\tilde{g}^I_{K,\alpha},\tilde{g}^I_{\bar{K},\alpha};\,K\ge 0$)
are defined by
\begin{equation}
\left\{
\begin{array}{l}
{\displaystyle
 \tilde{g}^I_{K,\alpha} \equiv \frac{1}{\sqrt{2(1+\delta_{K0})}}
    \left[ g^I_{K,\alpha} + (-1)^{I+K} g^I_{-K,\alpha} \right]
	\mbox{ for }\  r_y=+1,}\\
{\displaystyle
 \tilde{g}^I_{\bar{K},\alpha} \equiv \frac{1}{\sqrt{2(1+\delta_{K0})}}
    \left[ g^I_{K,\alpha} - (-1)^{I+K} g^I_{-K,\alpha} \right]
	\mbox{ for }\  r_y=-1,}
  \end{array} \right.
\label{eq:Ryg}
\end{equation}
and the modified projectors $(\tilde{P}^I_{MK},\tilde{P}^I_{M\bar{K}})$
are defined in the same way.
For the non-cranked mean-field state $|\Phi\rangle$,
which is totally $D_2$-symmetric, only the components compatible to
the representation $(r_x,r_y,r_z)=(+1,+1,+1)$ survive;
\begin{equation}
 |\Psi^I_{M,\alpha}\rangle =
  \sum_{K={\rm even}\ge 0}
   \tilde{g}^I_{K,\alpha}\, \tilde{P}^I_{MK} |\Phi\rangle.
\label{eq:nonCrproj}
\end{equation}

If the cranking frequency vector $\bm{\omega}$ is small
the first order perturbation theory can be applied for
the selfconsistent mean-field state $|\Phi(\bm{\omega})\rangle$
calculated with the routhian in Eq.~(\ref{eq:Hcrank});
\begin{equation}
 |\Phi(\bm{\omega})\rangle
 \approx |\Phi\rangle
 +\bm{\omega}\cdot \bm{C}|\Phi\rangle,
\label{eq:Crpert}
\end{equation}
where $\bm{C}=(C_x,C_y,C_z)$ 
are one-body operators defined with respect to $|\Phi\rangle$
and related to the so-called angle operators canonically conjugate
to $(J_x,J_y,J_z)$ in the RPA~\cite{RS80}.
If the total Hamiltonian in Eq.~(\ref{eq:Hcrank}) were approximated
by the mean-field Hamiltonian neglecting the effect of the residual interaction,
then the operator $C_i$ $(i=x,y,z)$ could be written in a simple well-known form:
\begin{equation}
 H \approx h=\sum_{\alpha}E_\alpha a^\dagger_\alpha a^{}_\alpha
 \quad\Rightarrow\quad
 C_i\approx
 \sum_{\alpha>\beta}\biggl[\frac{(J_i)_{\alpha\beta}}{E_\alpha+E_\beta}
  a^\dagger_\alpha a^\dagger_\beta - \mbox{h.c.}\biggr],
\label{eq:Crap}
\end{equation}
where $(a^\dagger_\alpha, a^{}_\alpha)$ are the creation and annihilation
operators of the quasiparticle, $E_\alpha$ is the quasiparticle energy,
and $(J_i)_{\alpha\beta}$ is the matrix element of the operator $J_i$
with respect to the quasiparticle states.
Note that the operators $(C_x,C_y,C_z)$ have
the same $D_2$-symmetry property as $(J_x,J_y,J_z)$ for
the totally $D_2$-symmetric state $|\Phi\rangle$.
Taking into account the fact that the operators $J_x$, $J_y$, and $J_z$
belong to the representation
$(r_x,r_y,r_z)=(+1,-1,-1)$, $(-1,+1,-1)$, and $(-1,-1,+1)$, respectively,
the projected state from the cranked mean-field state~(\ref{eq:Crpert})
can be classified into the four terms;
\begin{equation}
\begin{array}{l}
{\displaystyle
 |\Psi^I_{M,\alpha}\rangle =
 \sum_{K\ge 0}\left( \tilde{g}^I_{K,\alpha}\, \tilde{P}^I_{MK}
 +\tilde{g}^I_{\bar{K},\alpha}\, \tilde{P}^I_{M\bar{K}} \right)
  |\Phi(\bm{\omega})\rangle
  } \\
{\displaystyle
  \quad\qquad\  \approx
 \sum_{K={\rm even}\ge 0}
  \tilde{g}^I_{K,\alpha}\, \tilde{P}^I_{MK} |\Phi\rangle
 +\sum_{K={\rm odd}> 0}
  \omega_x\tilde{g}^I_{\bar{K},\alpha}\, \tilde{P}^I_{M\bar{K}} C_x |\Phi\rangle
  } \\
{\displaystyle
  \qquad\qquad
 +\sum_{K={\rm odd}> 0}
  \omega_y\tilde{g}^I_{K,\alpha}\, \tilde{P}^I_{MK} C_y|\Phi\rangle
 +\sum_{K={\rm even}\ge 0}
  \omega_z\tilde{g}^I_{\bar{K},\alpha}\, \tilde{P}^I_{M\bar{K}} C_z|\Phi\rangle
  }.
\end{array}
\label{eq:Crproj}
\end{equation}
Thus each component of the 3D cranking procedure in Eq.~(\ref{eq:Hcrank})
induces different time-odd terms that are classified according to
the $D_2$-symmetry quantum numbers $(r_x,r_y,r_z)$.
Moreover, three frequencies $(\omega_x,\omega_y,\omega_z)$ appear
in combination with the amplitudes $g^I_{K\alpha}$,
and so the change of frequencies can be absorbed
into the change of the amplitudes
when the  Hill-Wheeler equation~(\ref{eq:HW}) is solved (even the sign
of the frequencies does not matter).
Namely it has been shown that the result of the angular-momentum-projection
is independent of the infinitesimally small values of frequencies
$(\omega_x,\omega_y,\omega_z)$ within the first order perturbation theory.

In the case of axial symmetry around the $z$-axis the operator $C_z$ vanishes
and the remaining two terms $C_x|\Phi\rangle$ and $C_y|\Phi\rangle$ give
the same contribution (the $x$- and $y$-axes are equivalent).
Therefore there is only one extra term instead of three
included by the infinitesimal cranking.

In Ref.~\cite{TS12} it was found that the spectrum obtained by projection
from the cranked HFB state with vanishingly small frequency is
different from the one obtained by projection from the non-cranked HFB state.
Namely the projected spectrum is discontinuous in the zero frequency limit.
The reason is quite obvious from the argument above; the projected state
from the cranked HFB state~(\ref{eq:Crproj}) is always different
from the one projected from the non-cranked HFB state~(\ref{eq:nonCrproj})
even if the frequencies are vanishingly small but non-zero.

\section{Results of numerical calculation}
\label{sec:result}

In the present work we adopt the finite range Gogny force with
the D1S parameterization~\cite{D1S} as an effective interaction throughout;
the treatment of the Gogny force is the same as Ref.~\cite{STS15}.
Therefore there is no ambiguity about the Hamiltonian.
We apply the theoretical framework of the previous section
to a typical rare earth nuclei, $^{164}$Er,
where the low-lying gamma vibrational $2^+_\gamma$ state 
is observed at the excitation energy, 0.860 MeV.
The constrained HFB Hamiltonian is diagonalized
in the basis of the isotropic harmonic oscillator potential
with the frequency $\hbar\omega=41/A^{1/3}$ MeV, and the same basis
is utilized in the subsequent angular-momentum-projection calculation.
The size of the basis states is controlled by the oscillator quantum numbers;
all the states with $n_x+n_y+n_z \le N_{\rm osc}^{\rm max}$ are included.
$N_{\rm osc}^{\rm max}=10$ is used in the following calculations.
We do not intend to discuss very high-spin states and
we take $I_{\rm max}=40$ and $K_{\rm max}=30$.
The number of mesh points $(N_\alpha,N_\beta,N_\gamma)$
for the Euler angles $\Omega=(\alpha,\beta,\gamma)$
utilized in the numerical integration of the angular-momentum-projection
operator are chosen to be $N_\alpha=N_\gamma=62$ and $N_\beta=82$, which
is checked to be enough even for large triaxial deformations.
The canonical basis cut-off parameter is taken to be $10^{-6}$ as
in the all previous calculations, and the norm cut-off parameter in
the configuration-mixing (GCM) calculation is taken to be $10^{-10}$.
As for the values of frequencies $\bm{\omega}\equiv(\omega_x,\omega_y,\omega_z)$
in Eq.~(\ref{eq:Hcrank}) for the infinitesimal cranking,
appropriately small values should be chosen;
it should be small enough for the first order perturbation theory to be valid,
while it should not be too small so that the relative magnitude of
the induced time-odd components are well above the numerical accuracy.
We use 10 keV$/\hbar$ for the value of the infinitesimal frequencies,
which is small enough to guarantee the independence of the result
for excitation spectrum within about one keV for low-lying states.
 
Calculated mean-field parameters of the ground state of $^{164}$Er are
$Q=5.644$ b and $\beta_2=0.316$ for the quadrupole deformation,
which is axially symmetric,
and $\bar{\Delta}_\nu=0.846$ MeV and $\bar{\Delta}_\pi=0.859$ MeV
for the neutron and proton average pairing gaps, respectively.
Here $\bar{\Delta}$ is defined by
$\bar{\Delta}\equiv
-\bigl({\sum_{a>b}\Delta_{ab}\kappa^*_{ab}}\bigr)
/\bigl({\sum_{a>b}\kappa^*_{ab}}\bigr)$,
where $\kappa_{ab}$ is the abnormal density matrix
(the pairing tensor) and $\Delta_{ab}$ is the matrix element
of the pairing potential~\cite{RS80}.
These values are slightly different from those,
$\beta_2=0.311$, $\bar{\Delta}_\nu=0.874$ MeV
and $\bar{\Delta}_\pi=0.906$ MeV, in Ref.~\cite{STS15}
because of the different size of the model space
($N_{\rm osc}^{\rm max}=12$ in~\cite{STS15}).

\subsection{Effect of triaxial deformation}
\label{sec:Etridef}

There is no low-lying second $2^+$ state below 2.5 MeV
if the angular-momentum-projection calculation is performed
from the axially symmetric HFB state without cranking.
This is because the axially symmetric HFB state (without cranking)
has only $K=0$ components of the wave function; the $|K|=2$ components
are necessary to have low excitation energy for the second $2^+$ state,
which is $|K|=2$ mode while the first $2^+$ state is $K=0$ mode.
One way to obtain the second $2^+$ state is to construct
a coherent linear combination of many $|K|=2$ two quasiparticle states,
which is the way taken by the RPA method~\cite{RS80}.
Another easiest way is to include the $|K|=2$ components into the HFB state,
from which the projection is performed,
by explicitly breaking the axial symmetry.
This can be done for the quadrupole deformation by requiring the finite $\gamma$
deformation with the constraint field $-\lambda_{22}(Q_{22}+Q_{2-2})$,
where the c-number $\lambda_{22}$ is the Lagrange multiplier;
apparently it induces $|K|=2$ (and $=4,6,\cdots$) components
in the ground state wave function.
In this subsection we consider the effect neither of the infinitesimal cranking
nor of the configuration-mixing,
which will be investigated in the following subsections.

In the present work, we restrict, without loss of generality,
the triaxial deformation in the $0 \le \gamma \le 60^\circ$ sector
with the definition in Eq.~(\ref{eq:Qgamma}).
Namely, the lengths of inertia axes satisfy
\begin{equation}
\biggl\langle \sum_{a=1}^A\left(x^2\right)_a\biggr\rangle
 \le
\biggl\langle \sum_{a=1}^A\left(y^2\right)_a\biggr\rangle
 \le
\biggl\langle \sum_{a=1}^A\left(z^2\right)_a\biggr\rangle,
\label{eq:lxyz}
\end{equation}
where the expectation values are taken with respect to the HFB state,
and the first equality holds at $\gamma=0^\circ$
and the second at $\gamma=60^\circ$.

\begin{figure}[!htb]
\begin{center}
\includegraphics[width=80mm]{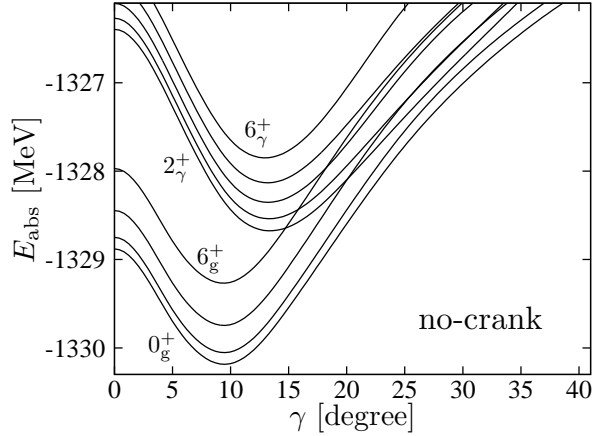}
\vspace*{-4mm}
\caption{
Absolute energy curves as functions of
the triaxial deformation parameter $\gamma$
calculated by the angular-momentum-projection from
the non-cranked HFB ground state in $^{164}$Er.
Those for the $0^+$, $2^+$, $4^+$, and $6^+$ states
of the g-band and
for the $2^+$, $3^+$, $4^+$, $5^+$, and $6^+$ states
of the $\gamma$-band are included.
}
\label{fig:ncAPES}
\end{center}
\end{figure}

First of all the calculated energies are shown
as functions of the triaxiality parameter $\gamma$ in Fig.~\ref{fig:ncAPES};
we include those of the $0^+$, $2^+$, $4^+$ and $6^+$ states
in the ground state band (g-band),
and of the $2^+$, $3^+$, $4^+$, $5^+$ and $6^+$ states
interpreted as the members of the gamma vibrational band ($\gamma$-band),
which are obtained by the angular-momentum-projection from
a single non-cranked HFB state with triaxiality $\gamma$
as in Eq.(\ref{eq:GCMgamNocm}).
The actual calculation is performed for ten $\gamma$ values,
$\gamma=1^\circ,5^\circ,10^\circ,\cdots,40^\circ,$ and $45^\circ$.
As in the case of the time-odd components induced by the cranking term,
the $|K| \ne 0$ components induced by the triaxial deformation
make the spectrum discontinuous at $\gamma=0$.
The spectrum obtained by the projection from the axially symmetric HFB state
is different from the one obtained from the HFB state with
vanishingly small but non-zero triaxial deformation.
We adopt $\gamma=1^\circ$ as a vanishingly small $\gamma$.
Utilizing the symmetry of energy $E(\gamma)=E(-\gamma)$,
the data are extended a few points to the negative values,
$\gamma=-10^\circ,-5^\circ,-1^\circ, 1^\circ,5^\circ,10^\circ,\cdots$,
and then continuous curves are generated by the cubic-spline interpolation
to make Fig.~\ref{fig:ncAPES}.
Therefore the data plotted at $\gamma=0$ in the figure are not
those obtained by setting $\gamma=0$
but the limiting values as $\gamma\rightarrow 0$.
As mentioned in Sec.~\ref{sec:ampcm},
the magnitude of the quadrupole moment $Q$ in Eq.~(\ref{eq:Qgamma}) is kept
at its ground state value in the calculation.
The deformation parameter $\beta_2$ slightly changes as a function of $\gamma$
because the mean-square-radius $\langle r^2 \rangle$
depends slightly on the triaxiality;
the amount of change in $\beta_2$ is small, within 2\%, however.
One can see that the ground state energy gains about 1.2 MeV by
the angular-momentum-projection at
the finite triaxial deformation of about $\gamma \approx 9.7^\circ$,
even if the minimum of the HFB energy is axially symmetric.
This is well-known for the angular-momentum-projection~\cite{HHR84,ETY00};
breaking the symmetry always introduces new degrees of freedom
and the associated correlation energy for its recovery
quite often defeats over the mean-field energy.
All curves for members of the g-band take minima at similar values,
$\gamma\approx 9.7^\circ$.
Therefore we can say that the members of the g-band have
almost the same deformation.
The same is true for members of the $\gamma$-band,
although the $\gamma$ values at minima, $\gamma\approx 13.6^\circ$,
are slightly larger than those of the g-band members.

\begin{figure}[!htb]
\begin{center}
\includegraphics[width=80mm]{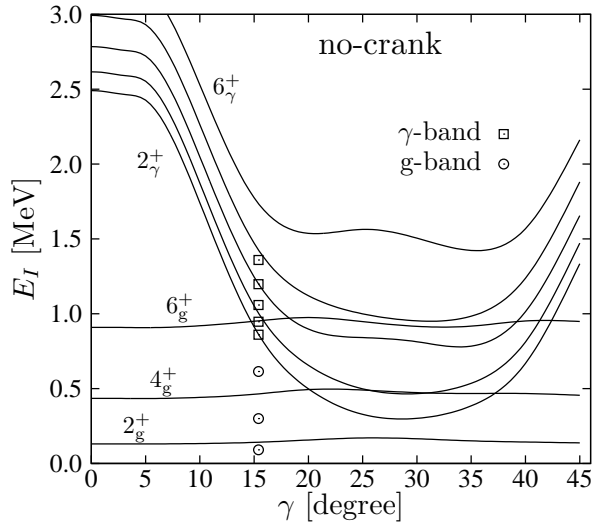}
\vspace*{-4mm}
\caption{
Excitation energy curves as functions of
the triaxial deformation parameter $\gamma$
calculated by the angular-momentum-projection from
the non-cranked HFB ground state in $^{164}$Er.
Those for the $2^+$, $4^+$ and $6^+$ states
of the g-band and
for the $2^+$, $3^+$, $4^+$, $5^+$ and $6^+$ states
of the $\gamma$-band are included.
Experimental data are also shown by symbols, open circles and squares,
at $\gamma=15.4^\circ$,
where the calculated excitation energy of $2^+_\gamma$ coincides
with the experimental one.
}
\label{fig:ncPES}
\end{center}
\end{figure}
 
Figure~\ref{fig:ncPES} shows the excitation energies
as functions of the triaxiality parameter $\gamma$
for the $2^+$, $4^+$ and $6^+$ states in the g-band,
and of the $2^+$, $3^+$, $4^+$, $5^+$ and $6^+$ states in the $\gamma$-band.
The same interpolation technique is used to draw Fig.~\ref{fig:ncPES}
as Fig.~\ref{fig:ncAPES}.
This result of the $\gamma$-dependence
can be compared with that of the asymmetric rotor model
with the irrotational moments of inertia in Ref.~\cite{DF58},
see e.g. Ref.~\cite{EG70} for a more complete figure of the spectrum.
Note that only the $\gamma$-dependence is meaningful
for this macroscopic rotor model.
The similarity is apparent: The excitation energies of the members of
the $\gamma$-band rapidly decrease as triaxiality increases
in the range $5^\circ \ltsim \gamma \ltsim 20^\circ$ and
the $2_{\gamma}^+$ energy crosses the $6_{\rm g}^+$ energy
at $\gamma\approx 15^\circ$.
Moreover, small bulges are observed at around $\gamma\approx 30^\circ$
in the $4^+_\gamma$ and $6^+_\gamma$ curves, which makes the energies of
the even-spin members higher than those of the odd-spin members leading to
the characteristic band structure of the wobbling band.
However, there are marked differences on the other hand:
The spectrum is symmetric with respect to the $\gamma=30^\circ$-axis
in the triaxial rotor model with the irrotational inertia but not
exactly symmetric in our microscopic angular-momentum-projection calculation,
and the bulges around $\gamma\approx 30^\circ$ 
in the $4^+_\gamma$ and $6^+_\gamma$ curves are not so pronounced
as in the case of the rotor model.  The most striking difference is that
the second $2^+$ state appears at around 2.5 MeV even with the small triaxiality
$\gamma \approx 1-5^\circ$ in the microscopic calculation;
however, the collectivity is not as high as experimentally observed.
At $\gamma\approx 9.7^\circ$, where the ground state energy takes minimum,
c.f. Fig.~\ref{fig:ncAPES}, the excitation energy of the second $2^+$ state
is about 1.9 MeV, while at $\gamma\approx 13.6^\circ$, where the absolute
energies of members of the $\gamma$-band take minima, it is about 1.1 MeV.
Compared with the experimental gamma vibrational energy, $0.860$ MeV,
the former value is considerably larger while the latter value is rather close.

\begin{figure}[!htb]
\begin{center}
\includegraphics[width=80mm]{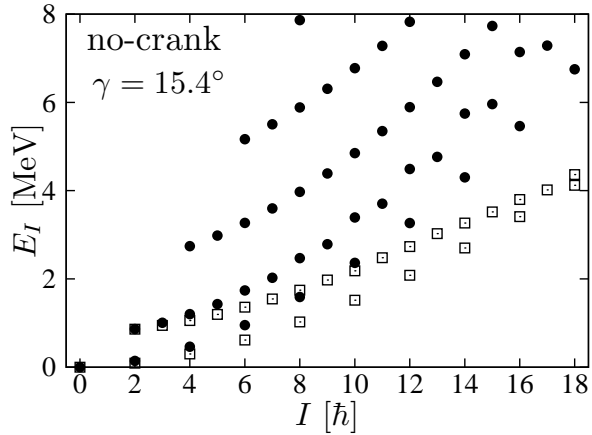}
\vspace*{-4mm}
\caption{
Energy spectrum calculated by the angular-momentum-projection from
the non-cranked HFB state at $\gamma=15.4^\circ$ in $^{164}$Er.
The experimental data for the g-band and the $\gamma$-band
are included as open squares.
}
\label{fig:nc15spt}
\end{center}
\end{figure}

The $\gamma$ value, at which the calculated $2^+_\gamma$ energy agrees
with the observed one, is $\gamma \approx 15.4^\circ$;
the experimental excitation energies are included in Fig.~\ref{fig:ncPES}
at this value.
We show in Fig.~\ref{fig:nc15spt} the calculated spectrum with this value,
$\gamma=15.4^\circ$, in comparison with experimental data.
Interestingly, the multiple band structure,
the one-, two-, and three-phonon excited bands starting from
$I=2^+_2$, $4^+_3$, and $6^+_4$ states, respectively,
are clearly seen with strong anharmonicity;
the excitation energy of the two(three)-phonon state, $4^+_3$($6^+_4$),
is considerably larger than the value which is two(three) times
that of the one-phonon state, $2^+_2$.  Moreover, the signature splitting
is observed in the high-spin part of the one-phonon band;
i.e., the odd-spin members are lower in energy than the even-spin member,
which is characteristic for the wobbling excitations.
This behavior is more strongly observed at larger value of the triaxiality
parameter $\gamma$, c.f. the next subsection.
However, the agreement of the calculated energies with the experimental data
is not satisfactory.
The moments of inertia of both the g-band and the $\gamma$-band are too small;
i.e., the energy spacings of the neighboring states in both bands are too large.
We need some improvements in order to obtain better agreement,
which will be considered in the next subsection.

A similar result has been reported by the triaxial projected shell model,
c.f. Fig.~1 of Ref.~\cite{SHS00}, which can be compared with
our result in Fig.~\ref{fig:ncPES}.  However,
one should be careful about the difference of the way of presentation.
In the works of the triaxial projected shell model,
the following deformed quadrupole potential is utilized:
\begin{equation}
-\frac{2}{3}\sqrt{\frac{4\pi}{5}}\frac{\hbar\omega_0}{b_0^2}
 \biggl[\epsilon\, Q_{20}+\epsilon'\frac{1}{\sqrt{2}}(Q_{22}+Q_{2-2})\biggr],
\label{eq:tpsmQpot}
\end{equation}
where the quantity $b_0$ is the oscillator length associated with
the oscillator frequency $\omega_0$.  In Ref.~\cite{SHS00} the excitation
energies are shown as functions of the parameter $\epsilon'$ with keeping
another parameter $\epsilon$ at the ground state value.
Our $\beta_2$ and $\gamma$ deformation parameters roughly correspond to
$\sqrt{\frac{16\pi}{45}}\sqrt{\epsilon^2+\epsilon'^2}$ and
$-\tan{(\epsilon'/\epsilon)}$, respectively.
Therefore, apart from the sign of $\gamma$,
which is irrelevant in the present context,
the $\beta_2$ value is increased when increasing $\epsilon'$
in the calculation of Ref.~\cite{SHS00}, while $\beta_2$ is kept constant
within 2\% in our calculation.
At first sight the result of Ref.~\cite{SHS00} is very similar to ours,
e.g., the excitation energies of members of the $\gamma$-band quickly
decrease as functions of the parameter $\epsilon'$,
although the agreement with the experimental data is much better than ours
when the value of $\epsilon'$ is appropriately chosen.
However, the value of triaxial deformation,
at which the calculated spectrum agrees with the experimental one,
seems quite different in Ref.~\cite{SHS00}; for $^{164}$Er nucleus,
$\epsilon=0.258$ and $\epsilon'=0.14$, which correspond to
$\beta_2\approx 0.310$ and $|\gamma|\approx 28.5^\circ$.
The value of $\beta_2$ is very similar to ours
$\beta_2\approx 0.316$ while that of $\gamma$ is much larger than
ours $\gamma \approx 15.4^\circ$.

On this difference, however, one has to be careful about
the definition of the triaxiality parameter:
The one utilized in Ref.~\cite{SHS00} is defined
with respect to the shape of the single-particle potential, $\gamma_{\rm pot}$,
while the one utilized in the present work is defined with respect to
the shape of the density distribution, $\gamma_{\rm den}$.
It has been discussed in Refs.~\cite{SSM08,SM84} that the difference between
these two parameters $\gamma_{\rm pot}$ and $\gamma_{\rm den}$ is rather
large for well-deformed nuclei.
Note that the $\gamma_{\rm pot}$ defined for the Nilsson potential is
still different from the one utilized in Eq.~(\ref{eq:tpsmQpot}),
$\gamma_{\rm pot}\equiv -\tan{(\epsilon'/\epsilon)}$.
For the harmonic oscillator potential model with the selfconsistent
deformation condition~\cite{BM75}, the transformation between
the deformation parameters in these different definitions can be done
easily, see Appendix of Ref.~\cite{SM84};
the values $\beta_{\rm pot}=0.310$ and $\gamma_{\rm pot}=28.5^\circ$
corresponds to $\beta_{\rm den}=0.334$ and $\gamma_{\rm den}=17.6^\circ$.
In this way the triaxiality $\gamma_{\rm den}$ is not very different
from our value $15.4^\circ$, although $\beta_2$ is a little bit larger.
Thus taking these differences into account the results
of Ref.~\cite{SHS00} and ours are rather consistent with each other.
However, the expected triaxial deformation
$\gamma_{\rm den} \approx 15^\circ-18^\circ$ is too large
if the $E2$ transition probability between the g-band and the $\gamma$-band
is investigated.

\begin{figure}[!htb]
\begin{center}
\includegraphics[width=80mm]{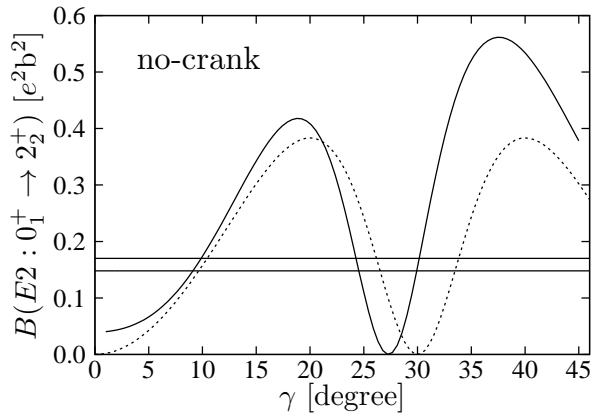}
\vspace*{-4mm}
\caption{
$E2$ transition probability from the ground state to the second $2^+$ state
as a function of the triaxial deformation parameter $\gamma$,
which is calculated by the angular-momentum-projection from
the non-cranked HFB state in $^{164}$Er (solid curve).
The dotted curve is the prediction of the asymmetric rotor model~\cite{DF58}
in Eq.~(\ref{eq:be2AR}) with $B=5.723$~[$e^2$b$^2$].
The experimentally measured values~\cite{RGH82},
0.148 and 0.170~[$e^2$b$^2$], are shown by two horizontal lines,
which are deduced from two different types of reactions.
}
\label{fig:ncBE2}
\end{center}
\end{figure}

Thus, we show in Fig.~\ref{fig:ncBE2} the $B(E2)$ value for the transition
from the $0^+$ ground state to the second $2^+$ state
as a function of $\gamma$ calculated by
the angular-momentum-projection from the non-cranked HFB state.
Note that we do not use any kind of effective charge
because the contributions of all nucleon are included.
In the figure the result of the asymmetric rotor model
with the irrotational inertia~\cite{DF58},
\begin{equation}
B(E2:0^+_1 \rightarrow 2^+_2)_{\rm AR}=\frac{B}{2}\,
  \biggl[1-\frac{3-2\sin^2(3\gamma)}{\sqrt{9-8\sin^2(3\gamma)}}\biggr]
  \ \mbox{[$e^2$b$^2$]},
\label{eq:be2AR}
\end{equation}
is also included as the dotted curve,
where $B$ is the in-band transition probability for the g-band,
$B(E2:0^+_1 \rightarrow 2^+_1)$ at $\gamma=0$.
The calculated value is $B=5.723$~[$e^2$b$^2$], which well corresponds to
the experimentally measured value, 5.81~[$e^2$b$^2$].
As is well-known, this $B(E2:0^+_1 \rightarrow 2^+_2)$ of the rotor model
rapidly increases with increasing the triaxiality, turns to decrease
at $\gamma \approx 20^\circ$, and vanishes at $\gamma=30^\circ$;
the $\gamma$-dependence is symmetric about $\gamma=30^\circ$-axis.
The rotor model wave function is represented with the Wigner-$D$ function
as $\sum_K f_{K,\alpha}^I {\cal D}_{MK}^I(\Omega)$,
where the amplitudes $f_{K,\alpha}^I$ corresponds to those properly normalized
in Eq.~(\ref{eq:normfNocm}), and the transition amplitude is evaluated as
\begin{equation}
 \langle 2^+_2 |\!| Q_2^{(E)} |\!| 0^+_1 \rangle
 \ \propto\  f_{K=0}^{2^+_2} \langle Q^{(E)}_{20} \rangle
        +2 f_{K=2}^{2^+_2} \langle Q^{(E)}_{22} \rangle
 =Q^{(E)} (f_{K=0}^{2^+_2} \cos\gamma - 2 f_{K=2}^{2^+_2} \sin\gamma),
\label{eq:Qe2}
\end{equation}
where $Q^{(E)}_{2\mu}$ is the electric quadrupole operator
(i.e., $e$ times the proton contribution) and
it is assumed that the deformation of neutrons and protons is the same.
For $\gamma < 20^\circ$ the $2^+_2$ state is almost purely $K=2$ mode
and the $K=0$ amplitude is almost negligible;
therefore the $B(E2)$ increases as $\propto\sin^2\gamma$
as $\gamma$ increases from 0
with the electric quadrupole moment $Q^{(E)}$ being fixed.
Further increasing $\gamma > 20^\circ$, however, the $K$-mixing quickly grows
and the two terms in Eq.~(\ref{eq:Qe2}) tends to cancel each other;
the exact cancellation occurs at $\gamma=30^\circ$ for the rotor model
with the irrotational inertia.
The microscopically calculated transition amplitude approximately satisfies
Eq.~(\ref{eq:Qe2}) and the resultant $B(E2)$ roughly follows
the trend of the rotor model, although it is not rigorously symmetric as
in the case of the excitation spectrum and vanishes at $\gamma\approx 27^\circ$.
Moreover the calculated value is considerably larger than that in
the rotor model at $\gamma \gtsim 30^\circ$.
The calculated $B(E2)$ values at small triaxiality,
$\gamma\approx 1^\circ-5^\circ$,
is about four to six times the single-particle unit,
which is two times the Weisskopf unit~\cite{BM75},
$B_{\rm s.p.}=2B_{\rm W}(E2)=0.0107$~[$e^2$b$^2$] in $^{164}$Er.
The experimentally measured value~\cite{RGH82}, 0.148 or 0.170~[$e^2$b$^2$],
is about fourteen to sixteen times the single-particle unit
and can be reproduced by the calculation with $\gamma\approx 10^\circ$.
Apparently if one employs the result with $\gamma \approx 15^\circ-18^\circ$
the $B(E2)$ value is overestimated by about a factor two to three as long as
the calculated quadrupole moment $Q^{(E)}$ is used,
which well reproduces the rotational $B(E2)$ values
inside the g-band assuming the axial symmetry ($\gamma=0$)
as it was demonstrated in our previous work~\cite{STS15}.

It is worthwhile mentioning that the calculated in-band $B(E2)$ value
for the g-band, $B(E2:0^+_1 \rightarrow 2^+_1)$,
as a function of $\gamma$ also well coincides
with the result of the asymmetric rotor model, which is obtained by changing
the first sign in Eq.~(\ref{eq:be2AR}) from $-$ to $+$~\cite{DF58},
although the deviation is non-negligible
for $\gamma \gtsim 27^\circ$ (not shown).

\subsection{Effect of infinitesimal cranking}
\label{sec:infc}

In the previous section it is found that the spectrum obtained
by the angular-momentum-projection from the non-cranked HFB state
is not very good in comparison with the experimental data.
The first problem is that the moment of inertia is too small,
which was already stressed in Ref.~\cite{TS12}.
Moreover, considerable triaxial deformation, $\gamma \gtsim 15^\circ$,
is necessary in order to reproduce the low-lying nature of the gamma vibration,
which is not justified from the energy minimization, c.f. Fig.~\ref{fig:ncAPES},
and from the $B(E2:0^+_1 \rightarrow 2^+_2)$ value;
the $B(E2)$ value for such triaxial deformation, $\gamma \gtsim 15^\circ$,
is too large compared with the measured value.
In order to achieve a better description of the rotational motion
for triaxial nuclei, we here study the effect of infinitesimal cranking,
which is explained in detail in Sec.~\ref{sec:crank}.

\begin{figure}[!htb]
\begin{center}
\includegraphics[width=125mm]{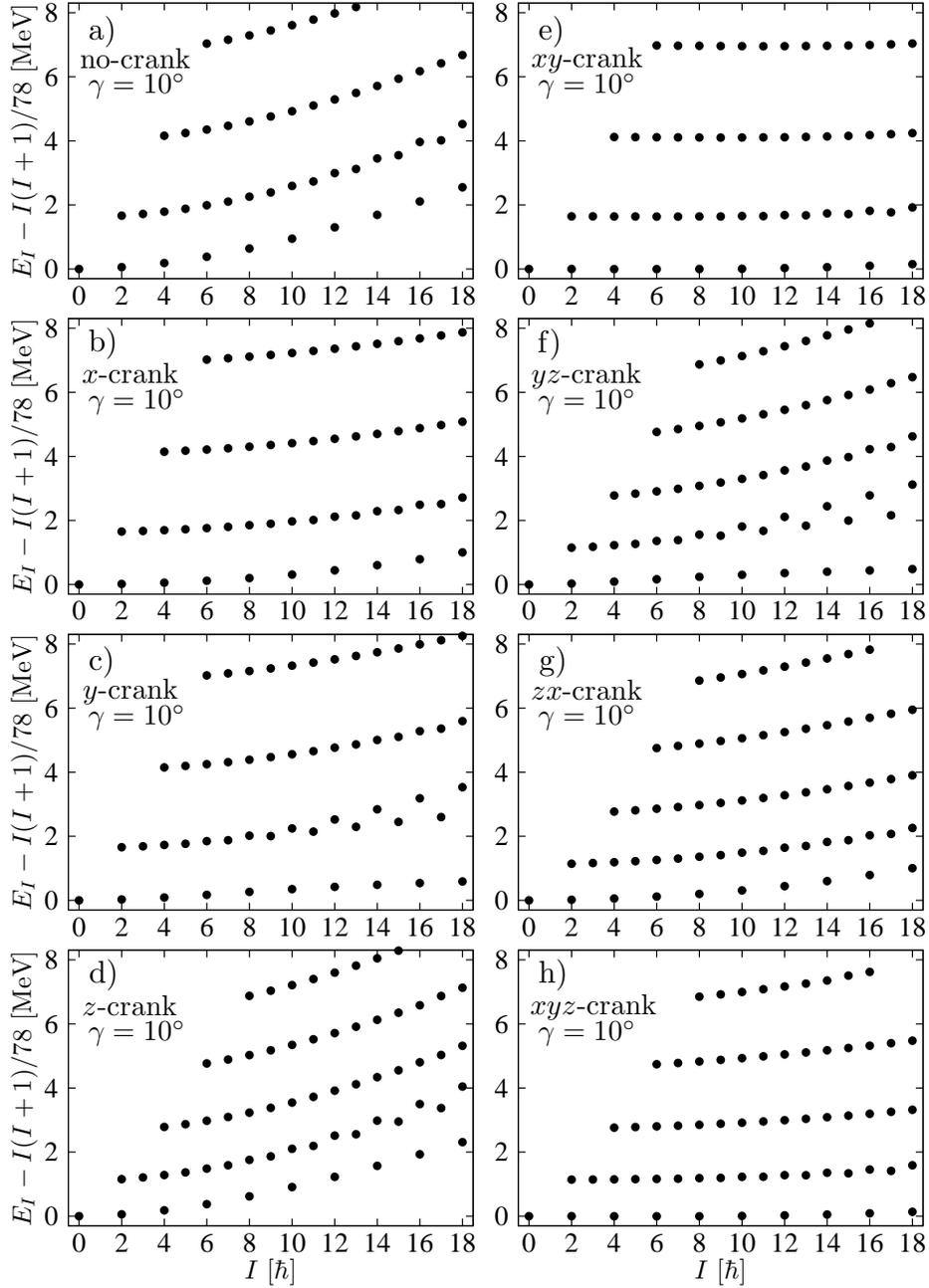}
\vspace*{-4mm}
\caption{
Energy spectrum calculated by the angular-momentum-projection
from the infinitesimally cranked HFB state
with triaxial deformation, $\gamma=10^\circ$, for $^{164}$Er.
The axis (axes) of cranking is (are) specified in each panel,
e.g., ``$xyz$-crank'' means that the infinitesimal cranking is performed
about all the $x,y$ and $z$-axes.  The non-cranked case is also
included as the panel~a).
}
\label{fig:gamma10}
\end{center}
\end{figure}

\begin{figure}[!htb]
\begin{center}
\includegraphics[width=125mm]{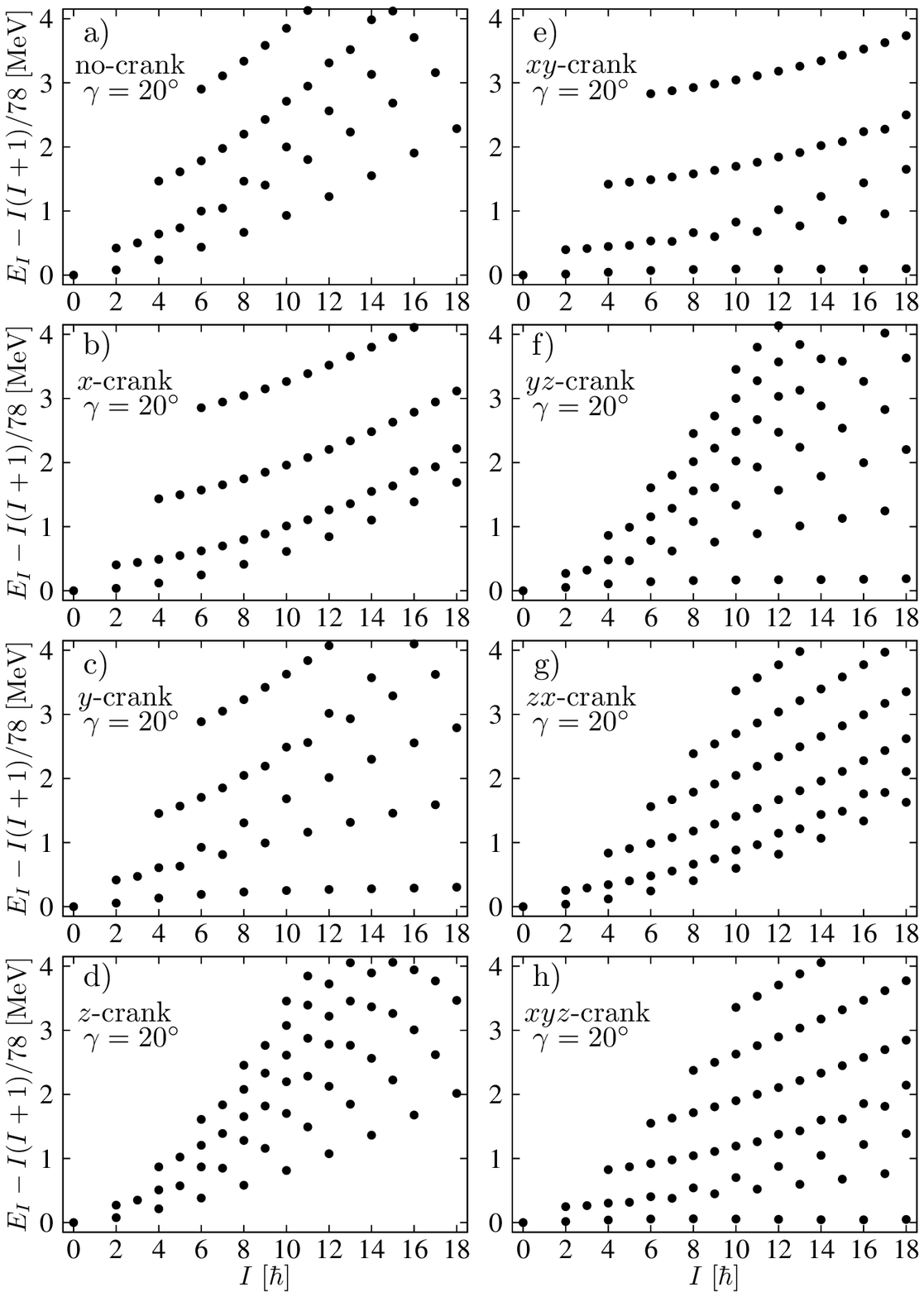}
\vspace*{-4mm}
\caption{
Energy spectrum calculated by the angular-momentum-projection
from the infinitesimally cranked HFB state
with triaxial deformation, $\gamma=20^\circ$, for $^{164}$Er.
The axis (axes) of cranking is (are) specified in each panel,
e.g., ``$xyz$-crank'' means that the infinitesimal cranking is performed
about all the $x,y$ and $z$-axes.  The non-cranked case is also
included as the panel~a).
}
\label{fig:gamma20}
\end{center}
\end{figure}

In Figs.~\ref{fig:gamma10} and~\ref{fig:gamma20} we show
how the spectrum changes for the triaxial deformation
with $\gamma=10^\circ$ and $\gamma=20^\circ$, respectively,
if the infinitesimally cranked HFB state is employed instead of
the non-cranked one for the angular-momentum-projection.
As it is discussed in Sec.~\ref{sec:crank}
there are three independent axes for cranking.
The results of all seven cases in addition to the non-cranked one are
included in the figures; the principal-axis cranking around
the $x$-, $y$- and $z$-axis, and the planer tilted-axis cranking around
the $xy$-, $yz$- and $zx$-axes, and finally the non-planer tilted-axis cranking
around all the $xyz$-axes.  In order to see each spectrum in more detail,
the reference rotational energy, $I(I+1)/(2{\cal J}_0)$, is subtracted
with ${\cal J}_0=39$~[$\hbar^2$/MeV], which roughly corresponds to
the average moment of inertia of
the experimentally observed g-band in $^{164}$Er.
In the case of smaller triaxiality $\gamma=10^\circ$
in Fig.~\ref{fig:gamma10}~a),
one can see nice multiple rotational bands without cranking,
as was already discussed in Fig.~\ref{fig:nc15spt}.
With the $x$- and $y$-axis cranking, Figs.~\ref{fig:gamma10}~b) and~c),
the slopes of the multiple bands decrease considerably, while
with the $z$-axis cranking, Fig.~\ref{fig:gamma10}~d),
the relative excitation energies of excited bands from the g-band decrease.
It can be seen that the excitation energy of the gamma vibration
is roughly 1.2 MeV if the cranking around the $z$-axis is performed,
while it is about 1.7 MeV without it for $\gamma=10^\circ$;
thus the considerable reduction of excitation energy is observed
by the $z$-axis cranking.
Comparing with the $x$- and $y$-axis cranking,
the increase of the moment of inertia is more or less the same
(or slightly larger with the $y$-axis cranking) for $\gamma=10^\circ$,
although the signature splitting of the one-phonon $\gamma$-band,
i.e., the splitting between the even-$I$ and odd-$I$ members, increases
by the $y$-axis cranking, while it decreases by the $x$-axis cranking.
The combination of cranking around two axes,
Figs.~\ref{fig:gamma10}~e)--g), gives more or less combined effects;
for example, with the $xy$-cranking the increase of the moments of inertia
for multiple bands is largest.  The results with the $yz$- and the $zx$-cranking
have similar multiple band structures but the signature splitting is
only apparent in the case of the $yz$-cranking.
With the cranking around all three axes in Fig.~\ref{fig:gamma10}~h),
the largest effect is observed and all multiple rotational bands are
approximately parallel and almost horizontal with smaller excitation energies
than those of the non-cranked case.

With larger triaxial deformation, $\gamma=20^\circ$, in Fig.~\ref{fig:gamma20},
the basic trend is similar, e.g., the reduction of the excitation energy of
the gamma vibration is largest with the $z$-axis cranking and
the large signature-splitting is induced by the $y$-axis cranking.
However, considerable differences from the case with
smaller triaxial deformation, $\gamma=10^\circ$, in Fig.~\ref{fig:gamma10}
are observed; the signature-splitting is much larger except for the cases
where the $x$-axis cranking is performed without the $y$-axis cranking
in Fig.~\ref{fig:gamma20}~b) and~g).
Generally increase of moments of inertia is observed for all multiple bands,
but the amounts of increase are somewhat different for each band
and those of the excited bands are not so large compared with
the case with smaller triaxiality, $\gamma=10^\circ$.
The only exception is the g-band, which is almost horizontal,
if the $y$-axis cranking is performed as is seen
in Figs.~\ref{fig:gamma20}~c), e), f), and~h); in the case of $\gamma=20^\circ$
the increase of inertia is largest with the $y$-axis cranking.
The signature-splitting is generally larger with larger triaxial deformation
as is seen even in the case without cranking;
compare Fig.~\ref{fig:gamma10}~a) and Fig.~\ref{fig:gamma20}~a).
The larger signature-splitting induced by the $y$- and $z$-axis cranking,
c.f. Figs.~\ref{fig:gamma20}~c), d), and~f), makes the band structure
as if it is composed of the even-$I$ and odd-$I$ sequences alternately,
which is characteristic for the wobbling rotational band.
With the cranking around all three axes in Fig.~\ref{fig:gamma20}~h),
nice multiple band structure appears with the wobbling-like structure
developing at higher-spin part.
The relation to the wobbling band will be briefly discussed
in the following subsection.

\begin{table}[ht]
\begin{center}
\begin{tabular}{cccccccccc}
\hline
$\gamma=10^\circ$ && no & $x$ & $y$ & $z$ &
 $xy$ & $yz$ & $zx$ & $xyz$ \cr
\hline
$E(2^+_1)$ [MeV] &&
 0.133 & 0.094 & 0.106 & 0.133 &
  0.076 & 0.105 & 0.094 & 0.076  \cr
$E(2^+_2)$ [MeV] &&
 1.742 & 1.732 & 1.735 & 1.237 &
  1.722 & 1.226 & 1.221 & 1.216  \cr
$B(E2)$ [$e^2$b$^2$] &&
 0.174 & 0.207 & 0.151 & 0.120 &
  0.182 & 0.100 & 0.158 & 0.135  \cr
\hline
\end{tabular}
\begin{tabular}{cccccccccc}
\hline
$\gamma=20^\circ$ && no & $x$ & $y$ & $z$ &
 $xy$ & $yz$ & $zx$ & $xyz$ \cr
\hline
$E(2^+_1)$ [MeV] &&
 0.157 & 0.114 & 0.132 & 0.154 &
  0.093 & 0.128 & 0.114 & 0.092  \cr
$E(2^+_2)$ [MeV] &&
 0.498 & 0.481 & 0.494 & 0.349 &
  0.474 & 0.347 & 0.330 & 0.324  \cr
$B(E2)$ [$e^2$b$^2$] &&
 0.409 & 0.675 & 0.316 & 0.199 &
  0.549 & 0.125 & 0.593 & 0.431  \cr
\hline
\end{tabular}
\end{center}
\caption{
The excitation energies of the first and second $2^+$ states
and the $B(E2:0^+_1 \rightarrow 2^+_2)$ value calculated by
the angular-momentum-projection from the infinitesimally cranked
HFB state around various axes; e.g., ``$xyz$'' means cranking around
all the three axes.  The upper and lower tables are for the triaxial
deformation, $\gamma=10^\circ$ and $20^\circ$, respectively.
}
\label{tab:infc}
\end{table}

In Table~\ref{tab:infc} the first and second excited $2^+$ energies
obtained by the angular-momentum-projection with the infinitesimal cranking
around various axes are summarized for $\gamma=10^\circ$ and $\gamma=20^\circ$.
The $B(E2)$ values of the transition from the ground state to the second
excited $2^+$ state are also included.
It is clear that the first $2^+$ state in the g-band
is lowered in energy by the $x$- and $y$-axis cranking,
while the second $2^+$ state in the $\gamma$-band
is lowered by the $z$-axis cranking, as was already discussed in relation to
Figs.~\ref{fig:gamma10} and~~\ref{fig:gamma20}.
As for the $B(E2)$ value, on the other hand,
it increases by the $x$-axis cranking,
while it decreases by the $y$- and $z$-axis cranking;
especially the $z$-axis cranking reduces the transition markedly,
and the simultaneous $yz$-cranking makes the $B(E2)$ value
about one third compared with the case without cranking
for the triaxiality $\gamma=20^\circ$.
In this way, it is interesting to see that the effects of cranking
around three independent axes are quite different
and the different combinations of rotation axes
considerably change the resultant angular-momentum-projected spectrum 
in the case of triaxial deformation.

\begin{figure}[!htb]
\begin{center}
\includegraphics[width=80mm]{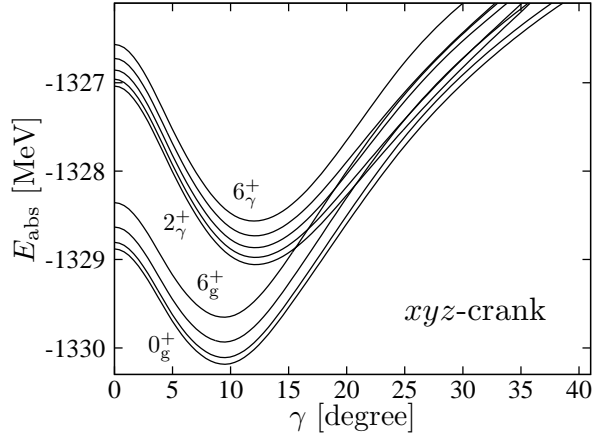}
\vspace*{-4mm}
\caption{
Absolute energy curves as functions of
the triaxial deformation parameter $\gamma$
calculated by the angular-momentum projection from
the $xyz$-cranked HFB ground state in $^{164}$Er.
Those for the $0^+$, $2^+$, $4^+$, and $6^+$ states
of the g-band and
for the $2^+$, $3^+$, $4^+$, $5^+$, and $6^+$ states
of the $\gamma$-band are included.
}
\label{fig:xyzAPES}
\end{center}
\end{figure}

\begin{figure}[!htb]
\begin{center}
\includegraphics[width=80mm]{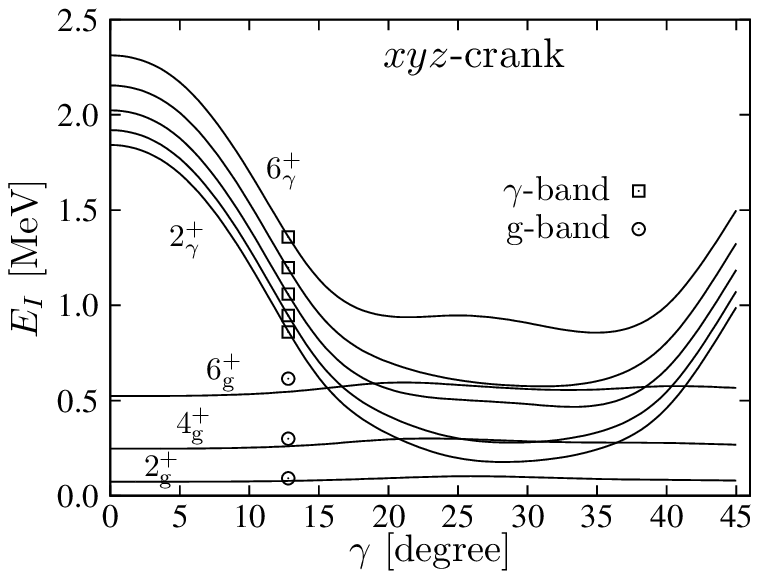}
\vspace*{-4mm}
\caption{
Excitation energy curves as functions of
the triaxial deformation parameter $\gamma$
calculated by the angular-momentum projection from
the $xyz$-cranked HFB ground state in $^{164}$Er.
Those for the $2^+$, $4^+$ and $6^+$ states
of the g-band and
for the $2^+$, $3^+$, $4^+$, $5^+$ and $6^+$ states
of the $\gamma$-band are included.
Experimental data are also shown by symbols, open circles and squares,
at $\gamma=12.8^\circ$,
where the calculated excitation energy of $2^+_\gamma$ coincides
with the experimental one.
}
\label{fig:xyzPES}
\end{center}
\end{figure}

\begin{figure}[!htb]
\begin{center}
\includegraphics[width=80mm]{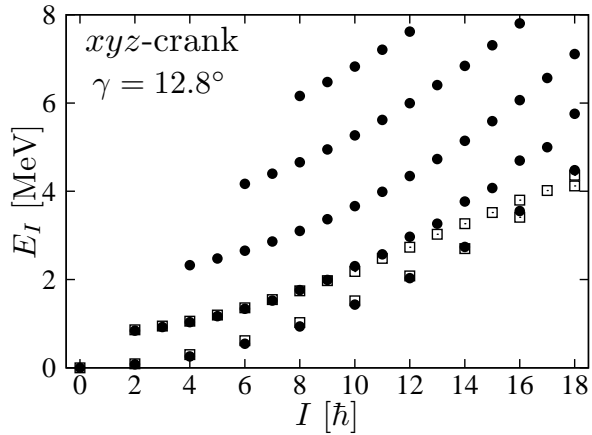}
\vspace*{-4mm}
\caption{
Energy spectrum calculated by the angular-momentum projection from
the $xyz$-cranked HFB state at $\gamma=12.8^\circ$ in $^{164}$Er.
The experimental data for the g-band and the $\gamma$-band
are included as open squares.
}
\label{fig:xyz13spt}
\end{center}
\end{figure}

From the variational point of view the cranking around all three axes
gives the best results in our theoretical framework.
With the infinitesimal cranking around all three axes,
i.e., the $xyz$-cranking, the calculated absolute energies
for members of the g-band and of the $\gamma$-band are shown
as functions of the triaxiality parameter $\gamma$ in Fig.~\ref{fig:xyzAPES}
as in the case without cranking in Fig.~\ref{fig:ncAPES},
and the calculated excitation energies are shown in Fig.~\ref{fig:xyzPES}
as in the case without cranking in Fig.~\ref{fig:ncPES}.
The same interpolation technique is used to draw
Figs.~\ref{fig:xyzAPES} and~\ref{fig:xyzPES} as
Figs.~\ref{fig:ncAPES} and~\ref{fig:ncPES}.
Comparing these two sets of figures,
the absolute $0^+$ ground state energy is very similar, while
the excitation energies of members of both the g-band and
the $\gamma$-band decrease considerably.  This means that
the moment of inertia is increased by the infinitesimal cranking
on one hand, and the excitation energy of the gamma vibration
is decreased on the other hand, which was already discussed
in relation to Figs.~\ref{fig:gamma10} and~\ref{fig:gamma20}.
The value of triaxiality $\gamma$,
which gives the minimum energy for the $0^+$ ground state,
is about $\gamma \approx 9.7^\circ$;
it is almost the same as in the case without cranking.
In contrast, the value which gives the minimum energy of
the $2^+$ gamma vibrational state is about $\gamma\approx 12.2^\circ$,
which is slightly smaller than the value $\gamma\approx 13.6^\circ$
without cranking.
In Fig.~\ref{fig:xyzPES} the experimental excitation energies
are included as symbols at $\gamma=12.8^\circ$,
where the second $2^+$ excitation energy is reproduced.
In contrast to the result shown in Fig.~\ref{fig:ncPES},
the agreement with the experimental excitation energies
at $\gamma=12.8^\circ$ is much better; clearly indicating
that the infinitesimal cranking improves the description of the $\gamma$-band.
In Fig.~\ref{fig:xyz13spt} the calculated spectrum using the $xyz$-cranked
HFB state with the triaxiality $\gamma=12.8^\circ$ is compared with
the experimental data as in the case without cranking in Fig.~\ref{fig:nc15spt}.
Apparently much better agreement with the experimental excitation energies
for both the g-band and the $\gamma$-band are obtained
with this infinitesimal cranking procedure.
It looks that the agreement of the $\gamma$-band becomes worse after
$I \ge 14$.  There is a reason for this: The band crossing occurs
for the experimental $\gamma$-band.  Namely, the $I \ge 14$ members
are interpreted as the states generated by exciting
the gamma vibration on the Stockholm-band (s-band),
the lowest two-quasineutrons-aligned band, not on the g-band.
We do not include the s-band configuration in the present work
and cannot describe the band crossing phenomenon in the $\gamma$-band.

\begin{figure}[!htb]
\begin{center}
\includegraphics[width=80mm]{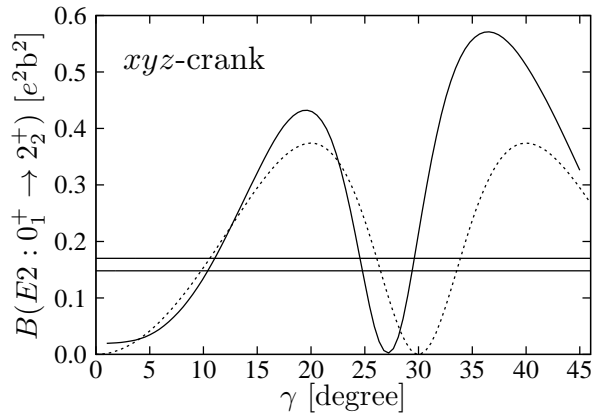}
\vspace*{-4mm}
\caption{
$E2$ transition probability from the ground state to the second $2^+$ state
as a function of the triaxial deformation parameter $\gamma$,
which is calculated by the angular-momentum-projection from
the infinitesimally cranked HFB state around all the $xyz$-axes
in $^{164}$Er (solid curve).
The dotted curve is the prediction of the asymmetric rotor model~\cite{DF58}
in Eq.~(\ref{eq:be2AR}) with $B=5.585$~[$e^2$b$^2$].
The experimentally measured values~\cite{RGH82},
0.148 and 0.170~[$e^2$b$^2$], are shown by two horizontal lines.
}
\label{fig:xyzBE2}
\end{center}
\end{figure}

In order to see the effect of the $xyz$-cranking on
the transition probability, we show in Fig.~\ref{fig:xyzBE2}
the calculated $B(E2)$ from the $0^+$ ground state
to the second excited $2^+$ state in the $\gamma$-band
as a function of the triaxiality $\gamma$.
The prediction of the triaxial rotor model in Eq.~(\ref{eq:be2AR})
is also included with the calculated value of $B=5.585$~[$e^2$b$^2$]
in the case of the $xyz$-cranking.
Compared to the result without cranking in Fig.~\ref{fig:ncBE2},
the general dependence of $B(E2)$ on $\gamma$ is similar, for example,
the $B(E2)$ values seem to vanish at $\gamma\approx 27^\circ$ in both cases.
Precisely speaking, however, the $B(E2)$ value does not vanish in the case with
the $xyz$-cranking; this is because the amplitude $g^I_{K,\alpha}$ is
complex in this case and the exact cancelation in Eq.~(\ref{eq:Qe2})
does not occur, although the imaginary part is very small so that
the actual $B(E2)$ value almost vanishes.
There are other marked differences:
The position of the lower peak moves to higher $\gamma$ value,
while that of the higher peak to lower $\gamma$ value,
and the $B(E2)$ values at both peaks are slightly larger
in the case with the $xyz$-cranking.
The largest difference is observed at lower $\gamma$ values,
$\gamma \ltsim 17^\circ$,
where the $B(E2)$ value calculated with the $xyz$-cranking
is smaller than that without cranking;
e.g., it is less than half in $\gamma < 5^\circ$.
The fact that the $B(E2)$ value with the $xyz$-cranking is larger
than that without cranking in the range,
$17.5^\circ \ltsim \gamma \ltsim 38^\circ$,
and is smaller otherwise can be seen also in Table~\ref{tab:infc};
the $B(E2)$ value reduces from 0.174 to 0.135 [$e^2$ b$^2$] at $\gamma=10^\circ$
while it increases from 0.409 to 0.431 [$e^2$ b$^2$] at $\gamma=20^\circ$.
As it will be discussed in the next subsection,
the expected $\gamma$ value is not so large, $\gamma \ltsim 15^\circ$,
and the effect of $xyz$-cranking appears to reduce the $B(E2)$ value,
which makes the agreement better with the experimentally measured value.

\subsection{Configuration mixing for triaxial deformation}
\label{sec:cmtrd}

Until the previous subsection, the triaxiality $\gamma$ is a parameter
and the results of the angular-momentum-projection calculation have been
presented as a function of it;
or in some cases the appropriate value of $\gamma$
is searched to reproduce the experimental data.
However, it should be determined theoretically,
or it should be treated properly to make theoretical predictions
independently of the experimental data.
In this subsection we show the result of configuration-mixing with respect to
the triaxiality parameter $\gamma$; namely the final wave function
is obtained by superposing the angular-momentum-projected states
as in Eq.~(\ref{eq:GCMgam}) in Sec.~\ref{sec:ampcm}.
Here five points have been employed for the $\gamma$ coordinate,
$\gamma=1^\circ$, $10^\circ$, $20^\circ$, $30^\circ$ and $40^\circ$.
We have checked that the excitation energies do not change within about 10 keV
by increasing the number of HFB states from five
in the range $0<\gamma \ltsim 45^\circ$ at least for the low-lying states.

\begin{figure}[!htb]
\begin{center}
\includegraphics[width=80mm]{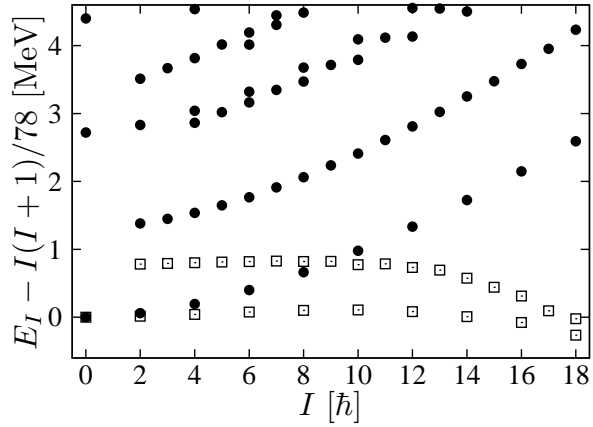}
\vspace*{-4mm}
\caption{
Energy spectrum calculated by the angular-momentum projected
configuration-mixing using the five non-cranked HFB states
with $\gamma=1^\circ$, $10^\circ$, $20^\circ$, $30^\circ$ and $40^\circ$
in $^{164}$Er.  The rotational energy $I(I+1)/78$ MeV is subtracted.
The experimental data for the g-band and the $\gamma$-band
are included as open squares.
}
\label{fig:ncspect}
\end{center}
\end{figure}

\begin{figure}[!htb]
\begin{center}
\includegraphics[width=80mm]{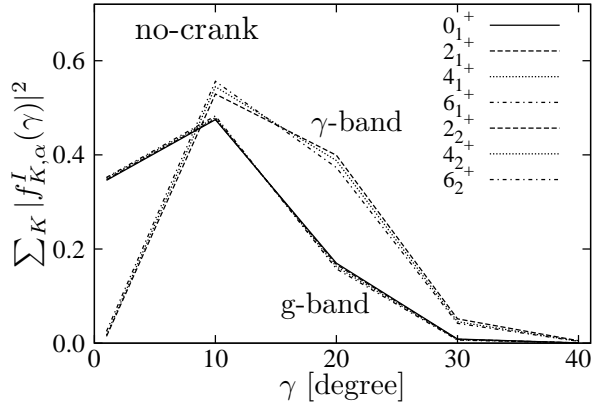}
\vspace*{-4mm}
\caption{
The probability distributions of
$0^+_1$, $2^+_1$, $4^+_1$ and $6^+_1$ states in the g-band
and $2^+_2$, $4^+_2$ and $6^+_2$ states in the $\gamma$-band
with respect to the triaxiality parameter $\gamma$
for the configuration-mixing calculation of Fig.~\ref{fig:ncspect}.
}
\label{fig:ncfamp}
\end{center}
\end{figure}

The resultant spectrum calculated by the configuration-mixing superposing
the five non-cranked triaxial HFB states after the angular-momentum-projection
is presented in Fig.~\ref{fig:ncspect}, where the reference rotational energy
is subtracted as in Figs.~\ref{fig:gamma10} and~\ref{fig:gamma20}.
In Fig.~\ref{fig:ncfamp} the probability distributions, Eq.~(\ref{eq:probf}),
for the selected members of both the g-band and the $\gamma$-band are shown.
The distributions for the members of each band are quite similar,
but those for the g-band and for the $\gamma$-band are different;
the average $\gamma$ value in the g-band, $\approx 9^\circ$, is smaller than
that in the $\gamma$-band, $\approx 15^\circ$, c.f. Table~\ref{tab:cm} below.
These average $\gamma$ values are close to those which give minima of
the absolute energies in Fig.~\ref{fig:ncAPES};
more precisely, $\gamma\approx 9.6^\circ$ for the g-band
and $\gamma\approx 13.6^\circ$ for the $\gamma$-band.
Thus the resultant triaxiality of the configuration-mixing is not so large,
although the distribution with respect to $\gamma$ is considerably broad.
From the spectrum shown in Fig.~\ref{fig:ncspect} one can see that
the moments of inertia for both the g-band and the $\gamma$-band are too small
and deviation from the experimental data increases rapidly at higher-spins.
This is because the non-cranked HFB states are employed. 
Moreover, the calculated excitation energy of the gamma vibration is too high,
$\approx 1.46$ MeV, compared with the experimental data,
$\approx 0.86$ MeV.
Thus the configuration-mixing does not help to improve the moments of inertia
nor the excitation energy of the gamma vibration.

It should be mentioned that qualitative change of the calculated spectrum
by the configuration-mixing is observed in Fig.~\ref{fig:ncspect}
in comparison with, e.g., Fig.~\ref{fig:gamma10}~a): New excited bands
appear at higher excitation energy, for example, an even-$I$ band starting
from the $0^+$ state at about 2.7 MeV.
In this calculation with non-cranked HFB states,
this new band starting from the $0^+$ state almost degenerates with
the band starting from the $4^+$ state at about 3.1 MeV (note that
the reference rotational energy is subtracted in Fig.~\ref{fig:ncspect}),
both of which are interpreted as ``two-phonon'' gamma vibrational bands.
In fact the excitation energies of band-head of these two bands are
almost twice of that of the (one-phonon) gamma vibrational state.
As it is well-known, there are two two-phonon gamma vibrational states
corresponding to the $K$ quantum numbers, $K=0$ and $K=4$.
As it was discussed in Ref.~\cite{SHS00}, no $0^+$ excited band appears
by the angular-momentum-projection from one triaxial mean-field state,
which is exactly the feature of the asymmetric rotor model with
the $(r_x,r_y,r_z)=(+1,+1,+1)$ $D_2$ symmetry~\cite{BM75};
this is also the case in our microscopic calculation.
With the configuration-mixing for the triaxial degree of freedom
we additionally obtain the $0^+$ excited band.  It may not be evident that
this $0^+$ excited band can be interpreted as the $K=0$ two-phonon $\gamma$-band.
We have compared the three $E2$ transition probabilities,
$B(E2:0^+_1 \rightarrow 2^+_2)$, $B(E2:2^+_2 \rightarrow 4^+_3)$ and
$B(E2:0^+_2 \rightarrow 2^+_2)$, which are
$B(E2:0^+_{\rm g.s.} \rightarrow 2^+_{\gamma})$,
$B(E2:2^+_{\gamma} \rightarrow 4^+_{\gamma\gamma})$ and
$B(E2:0^+_{\gamma\gamma} \rightarrow 2^+_{\gamma})$, respectively,
in an obvious notation,
and all coincide in the harmonic vibrational limit~\cite{BM82}.
The calculated values of these $B(E2)$'s are
0.222, 0.206 and 0.253 [$e^2$b$^2$], respectively;
they are indeed close with each other.
This result is in contrast to that of the other microscopic calculation
in Ref.~\cite{MM87}, where the calculated value
of $B(E2:0^+_{\gamma\gamma} \rightarrow 2^+_{\gamma})$
is considerably smaller than that
of $B(E2:2^+_{\gamma} \rightarrow 4^+_{\gamma\gamma})$.
It is interesting to investigate how such a difference appears
in the two microscopic calculations;
it is, however, out of scope in the present work.

\begin{figure}[!htb]
\begin{center}
\includegraphics[width=80mm]{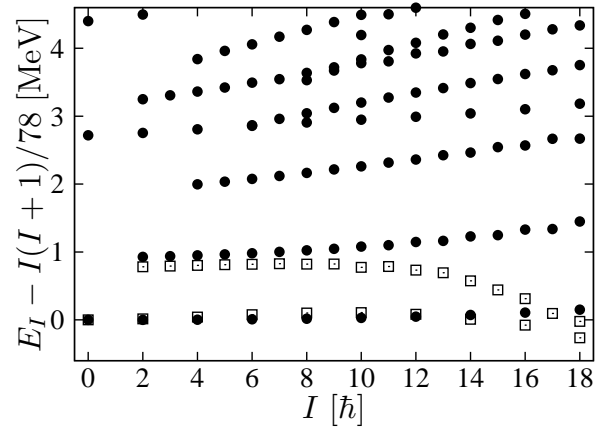}
\vspace*{-4mm}
\caption{
Energy spectrum calculated by the angular-momentum-projected
configuration-mixing using the five infinitesimally $xyz$-cranked HFB states
with $\gamma=1^\circ$, $10^\circ$, $20^\circ$, $30^\circ$ and $40^\circ$
in $^{164}$Er.  The rotational energy $I(I+1)/78$ MeV is subtracted.
The experimental data for the g-band and the $\gamma$-band
are included as open squares.
}
\label{fig:xyzspect}
\end{center}
\end{figure}

\begin{figure}[!htb]
\begin{center}
\includegraphics[width=80mm]{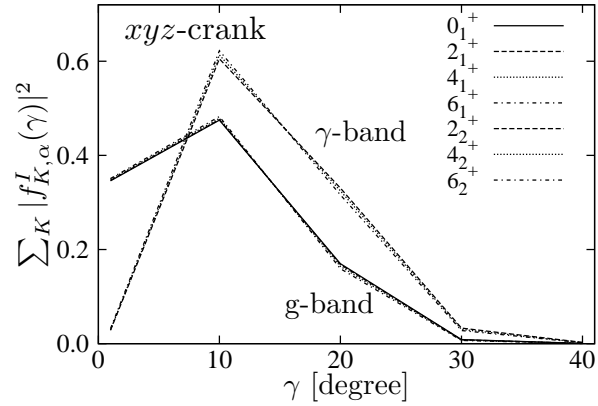}
\vspace*{-4mm}
\caption{
The probability distributions of
the $0^+_1$, $2^+_1$, $4^+_1$ and $6^+_1$ states in the g-band
and the $2^+_2$, $4^+_2$ and $6^+_2$ states in the $\gamma$-band
with respect to the triaxiality parameter $\gamma$
for the configuration-mixing calculation of Fig.~\ref{fig:xyzspect}.
}
\label{fig:xyzfamp}
\end{center}
\end{figure}

As for the moment of inertia and the excitation energy of
the gamma vibrational band, the infinitesimal cranking play an important role
as is discussed in the previous subsection.
We show the result of calculation by employing
the five $xyz$-cranked triaxial HFB states
with the same set of triaxiality parameters in Fig.~\ref{fig:xyzspect},
and the corresponding probability distribution in Fig.~\ref{fig:xyzfamp}.
Compared with the spectrum calculated without cranking in Fig.~\ref{fig:ncspect}
a great improvement has been achieved by the infinitesimal cranking.
The excitation energy of the gamma vibration, $\approx 1$ MeV, becomes close to
the experimental value.  The moments of inertia for both the g-band and
the $\gamma$-band are considerably increased, although they are still slightly
smaller at high-spin states.
The experimentally observed moments of inertia increase
as functions of spin, while the calculated inertias are
rather constant in the present work.
It was shown that the experimentally observed increasing feature
of the moment of inertia can be well reproduced
by superposing angular-momentum-projected configurations
with different values of the cranking frequency~\cite{STS15}.
It is, however, too heavy to perform the configuration-mixing
calculations taking into account both the cranking frequency
and the triaxial deformation at the same time.
We believe further improvements can be obtained with such calculations.
As for the probability distributions, those for the members of the g-band
are almost the same as in the case without cranking, while those for
the members of the $\gamma$-band slightly move to lower $\gamma$ values
and the widths of distribution become a little bit smaller
by the effect of the $xyz$-cranking, c.f. Table~\ref{tab:cm}.

Another interesting difference observed in Fig.~\ref{fig:xyzspect}
in comparison with Fig.~\ref{fig:ncspect} is that the degeneracy of
the two bands interpreted as the two-phonon $\gamma$-bands with $K=0$ and $K=4$
is resolved by the $xyz$-cranking.
The $0^+$ excited state at about 2.7 MeV keeps its excitation energy,
while the third $4^+$ state becomes lower in energy from about 3.1 to 2.3 MeV
(note that the reference rotational energy is subtracted
in Figs.~\ref{fig:ncspect} and~\ref{fig:xyzspect}).
With this effect the spectrum of the one- and two-phonon gamma vibrational states
becomes similar to that in other calculations, see e.g. Refs.~\cite{DH82,MM87};
namely energies of the two-phonon states are larger
than twice the energy of the one-phonon state,
i.e., large anharmonicity is observed,
and the $0^+$ state with $K=0$ lies higher than the $4^+$ state with $K=4$.
The three $B(E2)$ values,
$B(E2:0^+_1 \rightarrow 2^+_2)$, $B(E2:2^+_2 \rightarrow 4^+_3)$ and
$B(E2:0^+_2 \rightarrow 2^+_2)$ are
0.180, 0.164 and 0.176 [$e^2$b$^2$], respectively,
and so the interpretations of the $0^+$ and $4^+$ states as two-phonon
gamma vibrational states may be justified also with the $xyz$-cranking.

\begin{figure}[!htb]
\begin{center}
\includegraphics[width=80mm]{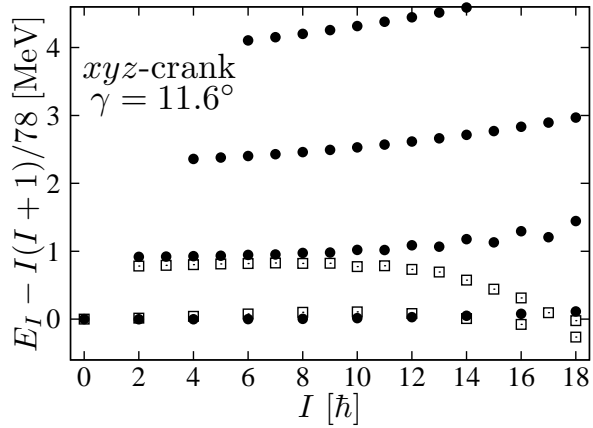}
\vspace*{-4mm}
\caption{
Energy spectrum calculated by the angular-momentum projection
from one $xyz$-cranked HFB state with $\gamma=11.6^\circ$ in $^{164}$Er.
The rotational energy $I(I+1)/78$ MeV is subtracted.
The experimental data for the g-band and the $\gamma$-band
are included as open squares.
}
\label{fig:xyzspcnm}
\end{center}
\end{figure}

In order to see the effect of superposing the five HFB states with different
triaxial deformations, we show in Fig.~\ref{fig:xyzspcnm} the calculated spectrum
with a single HFB state with $\gamma=11.6^\circ$, whose low-spin part
of the $\gamma$-band roughly coincides with the result of configuration-mixing
in Fig.~\ref{fig:xyzspect}.  For the g-band and the $\gamma$-band the resultant
spectra in the two figures are very similar; as for these two bands
the effect of the configuration-mixing is not very large.
However, other more excited bands are very different
if Figs.~\ref{fig:xyzspcnm} and~\ref{fig:xyzspect} are compared.
For example, excitation energies of the band starting from the $4^+_3$ state,
interpreted as one of the excited two-phonon $\gamma$-bands,
and of the band starting from the $6^+_4$, interpreted as a three-phonon band,
are considerably lower in Fig.~\ref{fig:xyzspect} than
those in Figs.~\ref{fig:xyzspcnm}.
Note that there are two almost completely degenerate $6^+$ states
at about 3.4 MeV, which are at about 2.9 MeV in Fig.~\ref{fig:xyzspect}
because of the subtraction of the rotational energy, $I(I+1)/78$ MeV.
Moreover, as already discussed, the band starting from the $0^+_2$ state,
which is interpreted as another two-phonon band with strong anharmonicity,
and the band from the $2^+_4$ state,
which is interpreted as one of other three-phonon bands, etc.
are missing in the angular-momentum-projection calculation
from a single HFB state in Fig.~\ref{fig:xyzspcnm}.
Therefore the effect of configuration-mixing is important
for the complete understanding of the multiple $\gamma$-bands,
although the experimental information of them,
especially for the higher excited bands, is still scarce.

\begin{table}[ht]
\begin{center}
\begin{tabular}{ccccccccc}
\hline
  && no & $x$ & $y$ & $yz$ & $zx$ & $xyz$ & exp. \cr
\hline
$E(2^+_1)$ [MeV] &&
 0.137 & 0.090 & 0.100 & 0.099 & 0.089 & 0.078 & 0.091 \cr
$E(2^+_2)$ [MeV] &&
 1.458 & 1.445 & 1.450 & 1.000 & 0.993 & 1.005 & 0.860 \cr
$B(E2)$ [$e^2$b$^2$] &&
 0.222 & 0.259 & 0.202 & 0.137 & 0.200 & 0.180 & 0.148/0.170 \cr
$\langle \gamma \rangle_{0^+_1}$ [$^\circ$] &&
  8.8 &  8.8 &  8.8 &  8.8 &  8.8 &  8.8 & $-$ \cr
$2({\mit\Delta}\gamma)_{0^+_1}$ [$^\circ$] &&
 13.8 & 13.8 & 13.8 & 13.8 & 13.8 & 13.8 & $-$ \cr
$\langle \gamma \rangle_{2^+_2}$ [$^\circ$] &&
 15.1 & 15.1 & 15.0 & 13.7 & 13.7 & 13.8 & $-$ \cr
$2({\mit\Delta}\gamma)_{2^+_2}$ [$^\circ$] &&
 12.8 & 12.8 & 12.8 & 12.1 & 12.1 & 12.1 & $-$ \cr
\hline
\end{tabular}
\end{center}
\caption{
The excitation energies of the first and second $2^+$ states
and the $B(E2:0^+_1 \rightarrow 2^+_2)$ value calculated by
the angular-momentum-projected configuration-mixing
with the infinitesimally cranked HFB states around various axes;
e.g., ``$yz$'' means cranking around the $y$- and $z$-axes.
The average triaxiality, $\langle \gamma \rangle$,
and two times the standard deviation,
$2{\mit\Delta}\gamma=2\sqrt{\langle (\gamma-\langle \gamma \rangle)^2 \rangle}$,
for the $0^+$ ground state and the $2^+$ gamma vibrational state
are also included.
The experimental data for the excitation energies and
the $B(E2)$ value~\cite{RGH82} are also tabulated at the last column.
}
\label{tab:cm}
\end{table}

Although our best result is the one employing the infinitesimal cranking
about all three axes, the $xyz$-cranking,
we have performed the configuration-mixing calculations for some other cases.
The results are summarized in Table~\ref{tab:cm}, where, for example,
``$yz$'' means that the five $yz$-cranked HFB states with the same set of
triaxial deformations,
$\gamma=1^\circ$, $10^\circ$, $20^\circ$, $30^\circ$ and $40^\circ$,
are superimposed.
In this table, the average $\gamma$ values calculated by the probability
distribution in Eq.~(\ref{eq:probf}),
\begin{equation}
 \langle \gamma \rangle_\alpha \equiv \int \gamma \, p^I_\alpha(\gamma)d\gamma
  \approx \sum_n \gamma_n \, p^I_\alpha(\gamma_n),
\label{eq:avegam}
\end{equation}
and the two times the standard deviation,
$2({\mit\Delta}\gamma)_{\alpha}\equiv
2\sqrt{\langle (\gamma-\langle \gamma \rangle_{\alpha})^2 \rangle}_\alpha$,
which roughly corresponds to the full width at half maximum,
are also included.
The basic feature is the same as that in the result of calculation
with using a single HFB state: The moment of inertia reflected by
the first $2^+$ energy is increased mainly by the $x$- and $y$-axis cranking,
and the second $2^+$ (i.e., the gamma vibrational) energy is lowered mainly
by the $z$-axis cranking.  The $B(E2)$ value is increased by the $x$-cranking,
while it is decreased by the $y$- and $z$-axis cranking.
These features are specific for the case where the triaxial deformation
is relatively small.  In fact the average $\gamma$ values are about $9^\circ$
for the g-band and about $13^\circ-15^\circ$ for the $\gamma$-band.
It may be worthwhile noticing that the infinitesimal cranking about
more than two-axes makes the average $\gamma$ value and
the width of distribution smaller for the $\gamma$-band,
while those for the g-band are not affected.

As for the rotational in-band $E2$ transitions for the g-band,
the result is similar to that of our previous axially symmetric calculation
in Ref.~\cite{STS15}, and agrees very well with the experimental data.
The reason is that the deformation parameter $\beta_2$ is very similar
and the triaxiality is rather small in the g-band.  Precisely speaking,
the $B(E2)$ value is about 2\% larger than that in Ref.~\cite{STS15}
at low-spins and the difference gradually increases up to about 10\%
at $I\approx 20$, where there are no experimental data available.
Thus the effect of configuration-mixing for the triaxial deformation
does not have a large impact for the $E2$ transitions inside the g-band.

\subsection{Relation to wobbling motion}
\label{sec:wob}

It was suggested in Ref.~\cite{MJ78} that a character change
from the gamma vibration to the wobbling motion is expected
in the high-spin continuation of the $\gamma$-band.
In fact, in the result of calculation with the larger triaxial deformation,
$\gamma=20^\circ$, the signature-splitting of the multi-phonon excited bands
becomes large at high-spins and the even-$I$ and odd-$I$ sequences alternately
compose a different type of multiple band structure from the one at low-spins,
as it is clearly seen in Figs.~\ref{fig:gamma20}~c), d), and~f).
We take the example of this calculation with $\gamma=20^\circ$
and briefly discuss the character change in the following, although
it does {\it not} correspond to the experimental situation of $^{164}$Er:
More complete discussion will be reported in a separate publication.

\begin{figure}[!htb]
\begin{center}
\includegraphics[width=80mm]{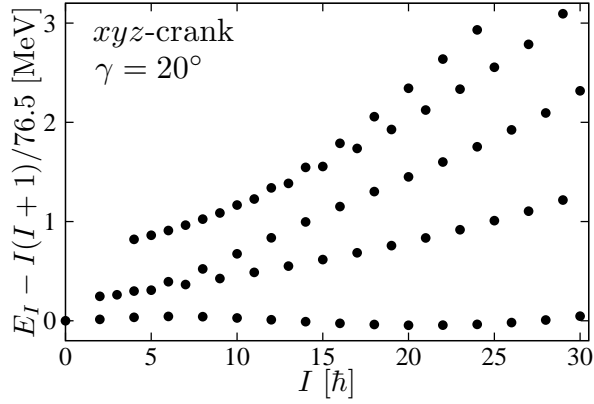}
\vspace*{-4mm}
\caption{
Energy spectrum calculated by angular-momentum-projection
from the one $xyz$-cranked HFB state with the triaxial deformation,
$\gamma=20^\circ$, for $^{164}$Er,
i.e., the same as Fig.~\ref{fig:gamma20}~h) but its higher-spin part is shown.
Only the five excited $|{\mit\Delta}I|=2$ bands from the lowest are shown.
}
\label{fig:wobex}
\end{center}
\end{figure}

We show the higher-spin continuation of the result of the $xyz$-cranking
in Figs.~\ref{fig:gamma20}~h), where a slightly
different rotational energy from that in Fig.~\ref{fig:gamma20}
is subtracted to make the yrast band as flat as possible.
It can be seen that a nice wobbling-like multiple band structure develops
at $I \gtsim 20$; the yrast band is composed of the even-$I$ states,
the first excited band is of the odd-$I$,
the second excited band is of the even-$I$, and so on.
The excitation energy of the first excited band increases
almost linearly as a function of spin as it is expected~\cite{BM75}.
It may be interesting to notice that the spectrum is not exactly
phonon-like in the sense that the excitation energy from the $(n-1)$-th
excited band to the $n$-th band decreases as $n$ increases ($n=1,2,3,\cdots$).

\begin{figure}[!htb]
\begin{center}
\includegraphics[width=80mm]{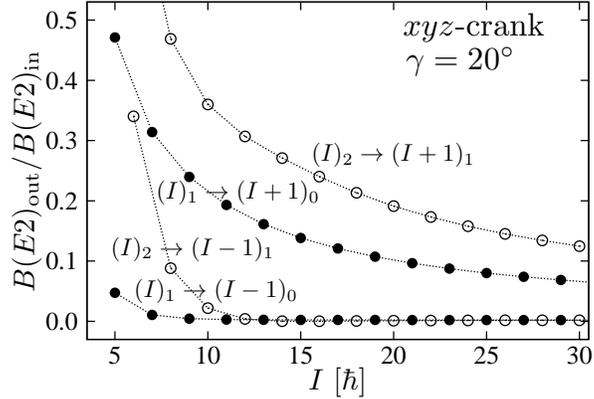}
\vspace*{-4mm}
\caption{
The $B(E2)$ ratio of the out-of-band to the in-band transitions,
$B(E2:I\rightarrow I\pm 1)/B(E2:I\rightarrow I-2)$, for the first and
the second excited bands in the wobbling-like structure in Fig.~\ref{fig:wobex}.
The filled (open) symbols denote the transitions from the first (second) band.
}
\label{fig:wobbe2r}
\end{center}
\end{figure}

The ratios of $E2$ transition probabilities, the out-of-band to the in-band,
$B(E2:I\rightarrow I\pm 1)/B(E2:I\rightarrow I-2)$, are shown as functions
of spin in Fig.~\ref{fig:wobbe2r} for the first and second excited bands.
The large out-of-band transition is one of the characteristic features
of the wobbling band, which is in fact employed as a guide
to identify it in experiments.
As is seen in the figure the out-of-band transitions are indeed very large
as expected.  It should be mentioned that there are two types of
out-of-band transitions, one is the $I \rightarrow I+1$ transition and
another is the $I \rightarrow I-1$ transition, c.f. Ref.~\cite{SM95},
and the former transitions are one or two orders of magnitude larger
in the present case.  One might wonder why;
the reason is that the main rotation axis is the $y$-axis in the present case.
The three nuclear moments of inertia behave like
those of the irrotational flow liquid,
and the largest inertia is that of the middle axis,
which is the $y$-axis as is mentioned in Eq.~(\ref{eq:lxyz}).
Namely the rotation of the present example is
of the so-called ``negative $\gamma$'' scheme,
where nucleus rotates mainly about the middle axis,
in contrast to the ``positive $\gamma$'' scheme
with the main rotation about the shortest axis.
Note that if the axis of rotation is chosen to be the $x$-axis,
as is usually done in the study of high-spin states~\cite{AL76},
the ``positive $\gamma$'' and ``negative $\gamma$'' rotation schemes
correspond to the triaxial shape with $0< \gamma < 60^\circ$ and
$-60^\circ < \gamma < 0$, respectively, from which the naming of them comes.

\begin{figure}[!htb]
\begin{center}
\includegraphics[width=80mm]{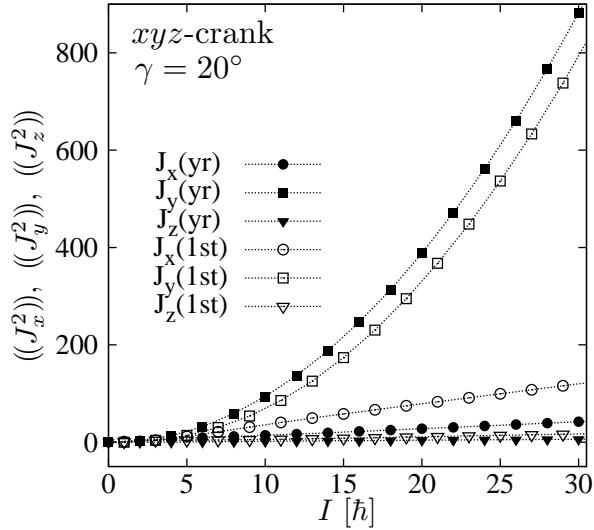}
\vspace*{-4mm}
\caption{
The ``expectation values'' of the squared $x,y,z$ components of
the angular momentum vector in the body-fixed frame,
which are defined by Eq.~(\ref{eq:exJJ}), for the yrast band (filled symbols)
and for the first excited band (open symbols) in Fig.~\ref{fig:wobex}.
}
\label{fig:wobjj}
\end{center}
\end{figure}

To confirm that the main rotation axis is the $y$-axis in the present example,
the expectation value of the angular momentum vector should be calculated
in the body-fixed frame, which is non trivial at all
for angular-momentum-projected wave functions.
In Ref.~\cite{GCS06} the result of such a calculation have been shown,
but unfortunately how to calculate is not explained.
In the present work, following the rotor model and using
the microscopically calculated normalized amplitudes
$\{f^I_{K,\alpha}\}$ in Eq.~(\ref{eq:normfNocm}),
we define the expectation value of the squared intrinsic component $J^2_i$
for the projected eigenstate $\alpha$ in the following way,
\begin{equation}
 (\!( J^2_i )\!)_\alpha
 \equiv \sum_{KK'} f^{I*}_{K,\alpha}\,
 \langle IK|J^2_i|IK'\rangle\,f^I_{K',\alpha},
\label{eq:exJJ}
\end{equation}
where $i=x,y,z$ denotes the axis of the body-fixed frame
specified by the deformed mean-field wave function $|\Phi\rangle$,
from which the projection is performed.
Needless to say, the purely algebraic quantity $\langle IK|J^2_i|IK'\rangle$
should be manipulated in the intrinsic frame.
The expectation values thus calculated
for the yrast band and for the first excited band
are shown in Fig.~\ref{fig:wobjj} as functions of spin.
Apparently the main rotation axis is the $y$-axis;
those of the $x$ and $z$ components increase almost linearly
as functions of spin, which is a typical behavior
for the angular momentum fluctuations.

\section{Summary and discussion}
\label{sec:summary}

In the present work, we have investigated the infinitesimal cranking
of the mean-field wave function in order to improve
the description of the collective rotational motion
by means of the angular-momentum-projected method.
For the triaxial deformation there are three axes for cranking.
Assuming the totally $D_2$ symmetric mean-field wave function
before the cranking, it is clarified what kind of different time-odd components
are induced into the wave function by the cranking about these three axes;
they are classified according to the $D_2$ symmetry quantum numbers
$(r_x,r_y,r_z)$.

Taking a nucleus $^{164}$Er as a typical example in the rare earth region,
we have firstly studied the spectrum and the $B(E2)$ values
by the angular-momentum-projection from a single HFB state
without cranking assuming the triaxial deformation.
The Gogny D1S force is employed as an effective interaction and
there is no adjustable parameter in the Hamiltonian.
As in the pioneering work of Ref.~\cite{SHS00}, the multiple $\gamma$-bands
appear in addition to the g-band by including the triaxial deformation
into the mean-field wave function.
The $\gamma$-dependences of the microscopically
calculated energy spectrum and $B(E2)$ values are similar to
those of the asymmetric rotor model with the irrotational moments of inertia
at least for $5^\circ \ltsim \gamma \ltsim 30^\circ$.
It has been found, however, that the moments of inertia for both the g-band
and the $\gamma$-band are too small compared with the experimental data,
and, moreover, rather large triaxial deformation is necessary to reproduce
the low-lying nature of the gamma vibration, with which the $B(E2)$ value
from the g-band to the $\gamma$-band is largely overestimated.
These problems of the result of the projection from a single HFB state
without cranking are shown to be greatly improved
if the infinitesimal cranking around all three axes is performed
for the mean-field wave function.
The effects of the infinitesimal cranking around the three axes are
quite different:  For $\gamma=10^\circ$ and $\gamma=20^\circ$
the $x$- and $y$-axis cranking mainly increase
the moments of inertia and the $z$-axis cranking
decreases the excitation energy of the gamma vibration;
furthermore, it was found that the $y$-axis cranking increases
the signature-splitting between the even-$I$ and the odd-$I$ sequences of
the $\gamma$-band, while the $x$-axis cranking decreases it.
With the infinitesimal cranking about all three axes a reasonable agreement
for both the spectrum and $B(E2)$ can be achieved with relatively small
triaxial deformation $\gamma\approx 12^\circ$.

In order to see what is the most probable triaxial deformation,
we have nextly performed the angular-momentum-projected configuration-mixing
calculation by superposing several HFB states
with different triaxial deformations.
The average $\gamma$ values for the g-band and the $\gamma$-band are
slightly different; they are about $9^\circ$ and $14^\circ$, respectively,
which are not so large and are comparable to the amplitude of
the zero-point oscillation estimated by the measured $B(E2)$ value~\cite{BM75}:
In fact the calculated width of the distributions
for the triaxial deformation is about $12^\circ - 14^\circ$.
Thus it does not conflict with the usual belief that the ground state
deformation is axially symmetric in the rare earth region.
With these calculated distributions for the triaxial deformation,
the resultant spectrum and the $B(E2)$ value agree reasonably well
with the experimental data, although the agreement is not perfect.
This is in contrast to the result of the RPA calculation~\cite{DH82}:
If the excitation energy is reproduced the $B(E2)$ value is largely
overestimated by a factor $2-4$.  Our calculation gives a correct magnitude
if the amplitude for the triaxial deformation is properly determined
by the configuration-mixing. It should be stressed that several new bands
appear at higher excitation energy by the configuration-mixing.
Especially, the $K=0$ two-phonon band, which is missing
in the projection calculation from a single HFB state, emerges
above the $K=4$ two-phonon band; this anharmonic pattern is very similar
to what was predicted by other calculations,
c.f. e.g. Refs.~\cite{DH82,MM85,MM86,MM87}.

Finally we have investigated the conjecture of Ref.~\cite{MJ78} that
the multiple $\gamma$-bands changes their character into the wobbling band.
By the hypothetical calculation with relatively large triaxial deformation,
$\gamma=20^\circ$, for $^{164}$Er,
it has been found that indeed the character-change occurs and the high-spin
part of the multiple $\gamma$-bands can be interpreted as the wobbling band.
The characteristic features of the calculated wobbling motion are studied:
The excitation energy of the wobbling-phonon almost linearly increases
as a function of spin
and the strong $I \rightarrow I+1$ out-of-band $E2$ transitions are predicted,
which are expected in the original work of the wobbling motion~\cite{BM75}
for the so-called ``negative $\gamma$'' rotation scheme~\cite{SM95}.
In experiment the wobbling band had been first observed in the odd nucleus,
$^{163}$Lu, in the rare earth region.  Interestingly enough, however,
the experimentally observed properties are opposite from what are predicted
in the present calculation.  The excitation energy decreases as spin increases,
which is now understood as the characteristic feature of the so-called
``transverse'' wobbling~\cite{FD14}, and only the $I \rightarrow I-1$
out-of-band $E2$ transitions are measured, which is characteristic
for the so-called ``positive $\gamma$'' rotation scheme~\cite{SM95}.
It should be pointed out that the possible occurrence of
the transverse wobbling was first pointed out in Ref.~\cite{SMM04},
where the effect of the aligned angular momentum of the odd particle
on the wobbling excitation energy was carefully examined.
We have also studied the wobbling motion in $^{163}$Lu
by the angular-momentum-projection method, and the preliminary
result was reported in Ref.~\cite{TSF14}, where the expected properties
for the case of $^{163}$Lu are reproduced, although the agreement
with the experimental data is not very satisfactory.
Thus, the wobbling motion appeared in the present hypothetical calculation
is somewhat different from what is observed in experiment.
We would like to notice, however, that the observed excitation energy
of the two-phonon wobbling state in $^{163}$Lu is smaller than twice the energy
of the one-phonon state, which roughly corresponds to what is seen
in the result of present calculation.
We have been investigating the wobbling motion in $^{163}$Lu
by performing similar calculation to the present work;
the result will be reported in a separate publication.

\section*{ACKNOWLEDGEMENTS}

Discussion with Prof.~Yang Sun,
when he visited Kyushu University, is greatly appreciated.
This work is supported in part
by Grant-in-Aid for Scientific Research (C) 
No.~25$\cdot$949 from Japan Society for the Promotion of Science.

\appendix*
\section{Expectation value in the intrinsic frame}

There is no concept of the intrinsic frame, or the body-fixed frame,
for the angular-momentum-projected wave function in Eq.~(\ref{eq:proj}).
Therefore, the expectation value of some operator in the intrinsic frame
is not an observable quantity and should be {\it defined} in some way.
In the text we have used the definition by Eq.~(\ref{eq:exJJ}) for
the squared component of the angular momentum vector,
but this definition is solely rely on the macroscopic rotor model and
can be applied only for the angular momentum operator
without any additional assumptions.
It is desirable to calculate the expectation value of an arbitrary operator
microscopically.
In this appendix we present some attempt following again the basic idea
of the rotor model;
the components of the spherical tensor operator in the body-fixed frame
are scalar and commute with the rotation operator.
For an arbitrary scalar observable ${\cal O}$ the expectation value
with respect to the projected wave function in Eq.~(\ref{eq:proj})
is written, just like for the Hamiltonian, as
\begin{equation}
 \langle \Psi^I_{M,\alpha}|{\cal O}| \Psi^I_{M,\alpha}\rangle
 = \sum_{KnK'n'} g^{I*}_{Kn,\alpha}\,
 \langle \Phi_n| {\cal O}P^I_{KK'}|\Phi_{n'} \rangle\,g^I_{K'n',\alpha}.
\label{eq:exOsc}
\end{equation}
Of course it does not depend on the $M$ quantum number.
However, if the observable ${\cal O}$ does not commute with the projector,
the right hand side is generally complex, so that one has to take
the real part or to symmetrize; thus we define the expectation value by
\begin{equation}
\begin{array}{ll}
 \langle\!\langle {\cal O} \rangle\!\rangle_\alpha
 & \equiv {\rm Re}\biggl(
  {\displaystyle \sum_{KnK'n'} g^{I*}_{Kn,\alpha}\,
 \langle \Phi_n|{\cal O} P^I_{KK'}|\Phi_{n'} \rangle
  \,g^I_{K'n',\alpha}}\biggr)
  \cr
 &= {\displaystyle \frac{1}{2}\sum_{KnK'n'} g^{I*}_{Kn,\alpha}\,
 \langle \Phi_n|({\cal O} P^I_{KK'}+ P^I_{KK'}{\cal O})|\Phi_{n'} \rangle
  \,g^I_{K'n',\alpha}},
\end{array}
\label{eq:exO}
\end{equation}
where because of this specific definition, we have used the notation,
$\langle\!\langle {\cal O} \rangle\!\rangle_\alpha$,
instead of an usual single bracket, and
the amplitudes $\{g^I_{Kn,\alpha}\}$ are assumed to be normalized.

\begin{figure}[!htb]
\begin{center}
\includegraphics[width=80mm]{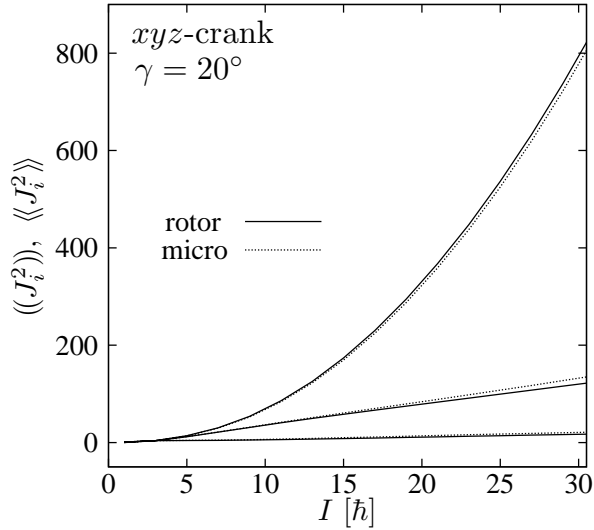}
\vspace*{-4mm}
\caption{
Comparison of the two definitions for the ``expectation values''
of the squared components of
the angular momentum vector in the body-fixed frame
for the first excited band in Fig.~\ref{fig:wobex},
which are calculated by Eq.~(\ref{eq:exJJ}) (solid lines)
and by Eq.~(\ref{eq:exJJm}) (dotted lines).
The solid lines are the same as those
with the open symbols in Fig.~\ref{fig:wobjj}.
}
\label{fig:wobjjrm}
\end{center}
\end{figure}

Then the expectation value of the squared intrinsic component $J^2_i$
can be microscopically calculated by
\begin{equation}
 \langle\!\langle J^2_i \rangle\!\rangle_\alpha
 \equiv {\rm Re}\biggl(\sum_{KK'} g^{I*}_{K,\alpha}\,
 \langle \Phi| J^2_i\, P^I_{KK'}|\Phi \rangle\,g^I_{K',\alpha}\biggr),
\label{eq:exJJm}
\end{equation}
where the configuration-mixing is neglected for simplicity
and the projection is performed from a single HFB state $|\Phi\rangle$.
This expression is a microscopic analog of Eq.~(\ref{eq:exJJ}),
in which the concept of the rotor model is fully employed.
In Fig.~\ref{fig:wobjjrm} we compare the results of two definitions,
Eqs.~(\ref{eq:exJJ}) and~(\ref{eq:exJJm}), for the first excited
band considered in Sec.~\ref{sec:wob}
(the result for the ground state band is similar).
The agreement of these two definitions is clear from the figure,
and the definition in Eq.~(\ref{eq:exJJm}) seems to be meaningful:
However, it is not always the case.
In fact the operators $J^2_i$ $(i=x,y,z)$ are not scalar but
a part of the second rank tensor, $X_{ij}\equiv\frac{1}{2}(J_i J_j+J_j J_i)$.
We have found that the expectation value of the non-diagonal part, e.g.,
$\langle\!\langle X_{yz} \rangle\!\rangle_\alpha$, depends on
the infinitesimal frequencies $(\omega_x,\omega_y,\omega_z)$,
and therefore can take arbitrary values
($\langle\!\langle X_{ij} \rangle\!\rangle_\alpha=0$ ($i \ne j$)
without cranking).
It can be confirmed that the diagonal part, 
$\langle\!\langle J^2_i \rangle\!\rangle_\alpha$, is independent of
these frequencies by using the $D_2$ symmetry;
the mean-field wave function before
the cranking is totally $D_2$ symmetric in the present case.
Therefore the definition in Eq.~(\ref{eq:exO}) does not always work.

\vspace*{10mm}


\end{document}